\documentclass{aa}  
\usepackage{graphicx}
\usepackage{amsmath}	
\usepackage{amssymb}	
\usepackage{relsize}
\usepackage{txfonts}
\usepackage[colorlinks=true,linkcolor=blue,allcolors=blue]{hyperref}%

\newcommand{\degree}{$^\circ$}

\begin{document}

   \title{Uncover 3D Dark Matter Distribution of the Milky Way by an Empirical Triaxial Orbit-Superposition Model: Method Validation}
   \subtitle{}

   \author{Ling Zhu$^1$\thanks{E-mail: lzhu@shao.ac.cn}, Xiang-Xiang Xue$^{2,5}$, Shude Mao$^3$, Chengqun Yang$^4$, Lan Zhang$^2$}

   \institute{
$^{1}$ Shanghai Astronomical Observatory, Chinese Academy of Sciences, 80 Nandan Road, Shanghai 200030, China \\
$^{2}$ National Astronomical Observatories, Chinese Academy of Sciences, Beijing 100101, China\\
$^{3}$ Department of Astronomy, Westlake University, Hangzhou, Zhejiang 310030, China\\
$^{4}$ School of Physics and Optoelectronic Engineering, Hainan University, 58 Renmin Avenue, Haikou, 570228, China\\
$^{5}$ Institute for Frontiers in Astronomy and Astrophysics, Beijing Normal University, Beijing 102206, China
}

   \date{Received; accepted}
   
   \titlerunning{an Empirical Triaxial Orbit-Superposition model}
\authorrunning{Zhu et al.}  
 
  \abstract
   {We introduce a novel dynamical model, named empirical triaxial orbit-superposition model, for the Milky Way halo. This model relies on minimal physical assumptions that the system is stationary, meaning the distribution function in 6D phase-space does not change when the stars orbit in the correct gravitational potential. We validate our method by applying it to mock datasets that mimic the observations of the Milky Way halo from LAMOST + Gaia with stars' 3D position and 3D velocity observed. By removing the stellar disk and substructures and correcting the selection function, we obtain a sample of smooth halo stars considered stationary and complete. We construct a gravitational potential including a highly flexible triaxial dark matter halo with adaptable parameters.
   Within each specified gravitational potential, we integrate orbits of these halo stars and build a model by superposing the orbits together taking the weights of stars derived from the selection function correction. 
   The goodness of the models are evaluated by comparing the density distributions as well as 3D velocity distributions numerically represented in the model to that in the data. The shape and radial density distribution of the underlying dark matter halo can be constrained well simultaneously.
  We applied it to three mock galaxies with different intrinsic shapes of their dark matter halos and achieved accurate recovery of the 3D dark matter density distributions for all. 
   }


   \maketitle
%

\section{Introduction}

The three-dimensional (3D) distribution of dark matter (DM) within the Milky Way (MW) is crucial to understanding its formation history. Despite extensive studies, there remain significant uncertainties regarding its total mass, density profile, and, in particular, its debated 3D shape. 

Streams, viewed as simple orbital pathways that trace the halo, played a significant role in constraining the shape of the MW DM halo. Several approaches have aimed to constrain the flattening of the DM using the two cold streams, Pal 5 and GD--1, which generally suggest a consistent result of $q=0.8-0.95$, assuming an axisymmetric halo \citep{Bovy2016ApJ...833...31B, Bowden2015MNRAS.449.1391B, Koposov2010ApJ...712..260K, kupper2015ApJ...803...80K, Malhan2019MNRAS.486.2995M}. These streams trace the inner halo at $r\lesssim$ 20 kpc. However, some other streams in similar regions indicate a preference for a more prolate halo. Recent research involving the Heimi stream has identified a mildly triaxial dark matter halo, with parameters $p = 1.013$ and $q = 1.204$ within the inner 20 kpc \citep{Woud2024A&A...691A.277W}. Another stream, M68 \citep{Palau2019MNRAS.488.1535P}, might also favour a prolate halo. The combination of data from NGC 3201, M68, and Palomar 5 fits well with a halo axis ratio $q=1.06$. \citep{Palau2023MNRAS.524.2124P}.

In the outer halo, the Sagittarius stream, being the most notable structure, is analysed using a variety of methods, which results in different outcomes with $q$ ranges from 0.44 to 1.3 \citet{ Pani2022arXiv221014983P, Law2010ApJ...714..229L, Johnston2005ApJ...619..800J, Helmi2004ApJ...610L..97H, Ibata2001ApJ...551..294I}. 
These inconsistent results could be mainly due to different parameterizations of the potential and treatment of the influence of the LMC on the outer portion of the stream \citep{Vera2013ApJ...773L...4V}.
Recent research shows that when modelling the Sagittarius stream in a dynamic potential, acknowledging our Galaxy's motion towards the massive in-falling LMC, the shape of the dark matter halo is not strongly constrained by the Sagittarius stream \citep{Vasiliev2021MNRAS.501.2279V}.
 
Conversely, halo stars--with full 6D phase-space information observed (3D position and 3D velocities)--span the entire space and, statistically, should exhibit stronger power in constraining the underlying DM distribution. However, these stars exist with complex orbital dynamics, necessitating accurate dynamical modeling. The current dynamical models applied to the MW halo usually impose ad hoc assumptions that might bias the results.

A commonly used dynamical model for the MW halo is the Jeans model, which often assumes either a spherical or axisymmetric configuration. Spherical Jeans models, which utilise giant stars in the halo extending to large radii $r \sim 100$ kpc, provide strong constraints on the MW halo mass \citep[e.g.,][]{Xue2008ApJ...684.1143X, Bird2022MNRAS.516..731B}, but do not restrict its shape by definition. The axisymmetric assumptions in the Jeans model generally yield mass estimates for the MW that are consistent with those of the spherical model (see review \citet{Wang2020SCPMA..6309801W}). However, the results diverge with respect to the DM halo's shape. 
Using RR Lyrae stars, research by \citet{Loebman2014ApJ...794..151L} identified an oblate halo with $q = 0.4 \pm 0.1$ at $r<20$ kpc, but \citet{Wegg2019MNRAS.485.3296W} found a nearly spherical halo with $q =1.00 \pm 0.09$ at $5<r<20$ kpc. The alignment of the velocity ellipsoid in the traditional axisymmetric Jeans model differs from observations of MW halo stars, potentially leading to bias. Recently, an axisymmetric Jeans model that allows for a generic shape and radial orientation of the velocity ellipsoid has been developed \citep{Nitschai2020MNRAS.494.6001N} and applied to MW BHB and K-giant stars, resulting in a near spherical DM halo with $q = 0.9-0.95$ \citep{Zhang2025}. Nevertheless, Jeans models are still limited in axisymmetric assumption and do not use the full 6D phase-space information, as they often only fit the first and second velocity moments and assume a specific tracer density.

To effectively use the 6D phase space data, distribution functions (DF) based dynamical models have been developed and implemented for the MW. A DF-based dynamical model applied to globular clusters (GCs) by \citep{Posti2019A&A...621A..56P} identifies a prolate halo with $q = 1.30 \pm 0.25$. In contrast, using DF-based axisymmetric models on halo stars reveals an oblate halo with $q > 0.963$ within $r<30$ kpc, as determined by RR Lyrae stars \citep{Hattori2021MNRAS.508.5468H, Li2022MNRAS.510.4706L}. These models presume an analytically assumed DF. A new approach named the Empirical Distribution Function has been proposed recently; it uses the observed 6D data of the tracers to construct the DF through appropriate smoothing of the observational data based on theoretical insights, making the DF data-driven with fewer inherent assumptions. Nevertheless, this model is restricted to spherical symmetry \citep{Han2016MNRAS.456.1017H, Li2024arXiv240811414L}. Recently, a novel method was developed to determine the gravitational potential of the MW using 6D data, using deep learning tools \citep{Green2023ApJ...942...26G}, making minimal physical assumptions that the system is stationary. This method learns the DF from the data and does not restrict the gravitational potential; however, it has not been applied to the MW halo yet, as it may require $\times 10^5$ data points to construct the model.

Currently, there is no triaxial dynamical model that properly represents the MW halo. However, triaxial particle-based Made-to-Measure \citep{Portail2017MNRAS.465.1621P, Long2013MNRAS.428.3478L} and orbit-based Schwarzschild models \citep{Khobulge2024arXiv241118182K} are widely used for nearby galaxies and the MW bulge. 
In such models, the DF of the tracer is numerically constructed by superposition of stellar orbits. 
For nearby galaxies, the observational information is integrated along the line-of-sight, without resolution of single stars; the true stellar orbit distribution is unknown. 
In the traditional Schwarzschild orbit-superposition method, a comprehensive orbit library is created based on theoretical assumptions, and orbit weights are determined by fitting to the observational data, i.e., line-of-sight velocity distribution across the 2D projected plane \citep[e.g.,][]{vdB2008, Zhu2018a, Thater2022A&A...667A..51T, Tah2024MNRAS.534..861T, Vasiliev.2015}. The limited observational data restrict the precision of the DF composed of stellar orbits and the ability to determine the DM shape of nearby galaxies.
In the MW, we have measurements of the 6D phase space for individual stars. However, the bulge areas are significantly affected by dust extinction, leading to incomplete observations. Analytical models continue to be employed to represent the stellar density distribution and used as model input. Libraries of orbits based on theoretical assumptions, along with the determination of orbit weights through data fitting, remain necessary to model the MW bulge \citep{Khobulge2024arXiv241118182K, Portail2017MNRAS.465.1621P, Long2013MNRAS.428.3478L, kho2024arXiv241115062K} and disk \citep{Khodisk2024arXiv241116866K, Bovy2018MNRAS.473.2288B}. 

In the MW halo, using data from LAMOST and Gaia, a large sample of K-giant stars has been constructed, with distances measured to an accuracy of within $\lesssim 15\%$, line-of-sight velocities with a typical error of $\sim 10$ km/s, and accurate proper motions with a typical error of $\sim 30$ km/s for tangential velocities. These stars cover a substantial portion of the halo, especially in the northern hemisphere, and the selection function is accurately calibrated for volumes up to 50 kpc \citep{Liu2017RAA....17...96L, Yang2022AJ....164..241Y}. The DF function of the stellar halo is thus observationally known. 

In this paper, we introduce an empirical triaxial orbit-superposition model for the Milky Way halo. Comparing to traditional Made-to-Measure and Schwarzschild method, we do not theoretically sample the orbit/particle libraries or determine the particle/orbit weights by fitting the model to data. Instead, we integrate the orbits of stars with 6D phase-space information observed and take the weights of stars from correction of selection function, i.e., the orbit library and orbit weights are entirely determined from the observations.
We take the minimum physical assumption that the stellar halo is stationary, meaning the DF of stars integrated in a correct potential matches that of stars directly observed instantaneously, our model results will thus be highly data driven. However, this method requires high-quality observations and is confined to the Milky Way only.

We apply the method to mock datasets that mimic the LAMOST and Gaia observations. Constrained by halo stars in 6D phase space, we show that the 3D shape and the radial density distribution of the Milky Way DM halo can be determined simultaneously. 
The paper is organised in the following way: we describe the details of mock data created from Auriga simulations in Section 2, we introduce the method and its application to the mock data in Section 3, we show the results in Section 4 and discuss them in Section 5. We conclude in Section 6.

\begin{figure*}
\centering\includegraphics[width=18cm]{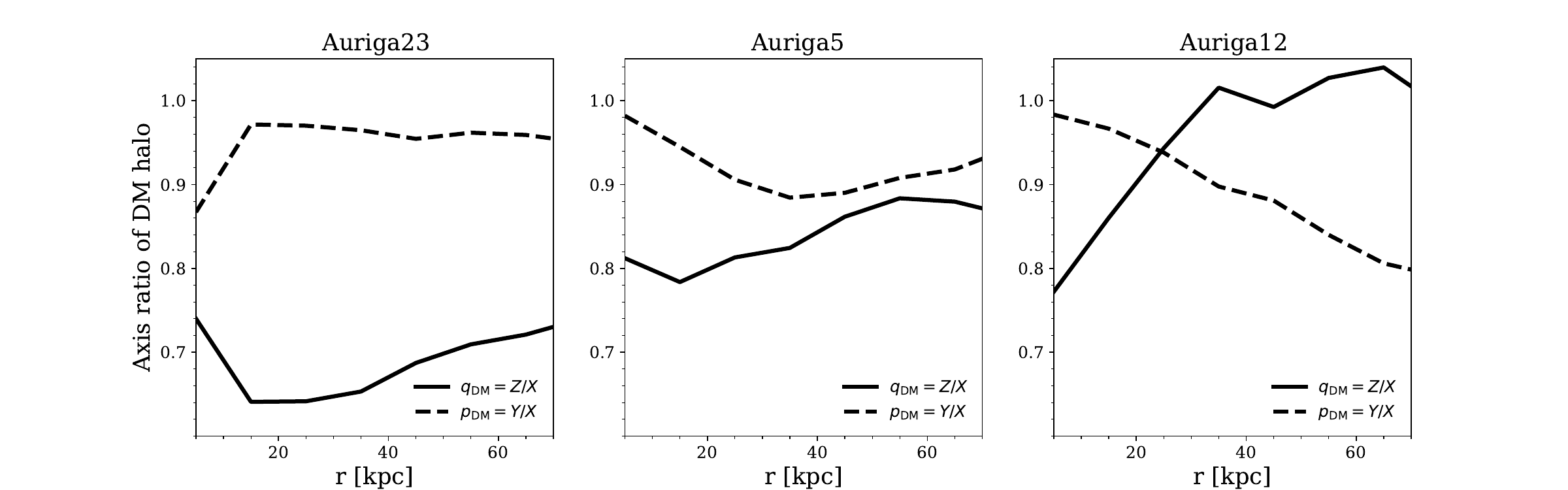}
\caption{The shapes of DM halos in Auriga 23, 5, 12, measured in the coordinate system where Z axis is perpendicular to the stellar disk, X and Y are the long and short axis of the DM halo in the disk plane. The DM halos of Auriga 23 and 5 are oblate aligned with the stellar disk ($p_{\rm DM}>q_{\rm DM}$) and vary little with radius. The DM shape of Auriga 12 varies from oblate aligned with the stellar disk ($p_{\rm DM} \sim 1$ and $p_{\rm DM}>q_{\rm DM}$) in the inner regions to be vertically aligned with the disk ($p_{\rm DM}<q_{\rm DM}$ and $q_{\rm DM} \sim 1$) at $r\gtrsim 20$ kpc.
}
\label{fig:vpq3_true}
\end{figure*}

\begin{figure*}
\centering\includegraphics[width=8.5cm]{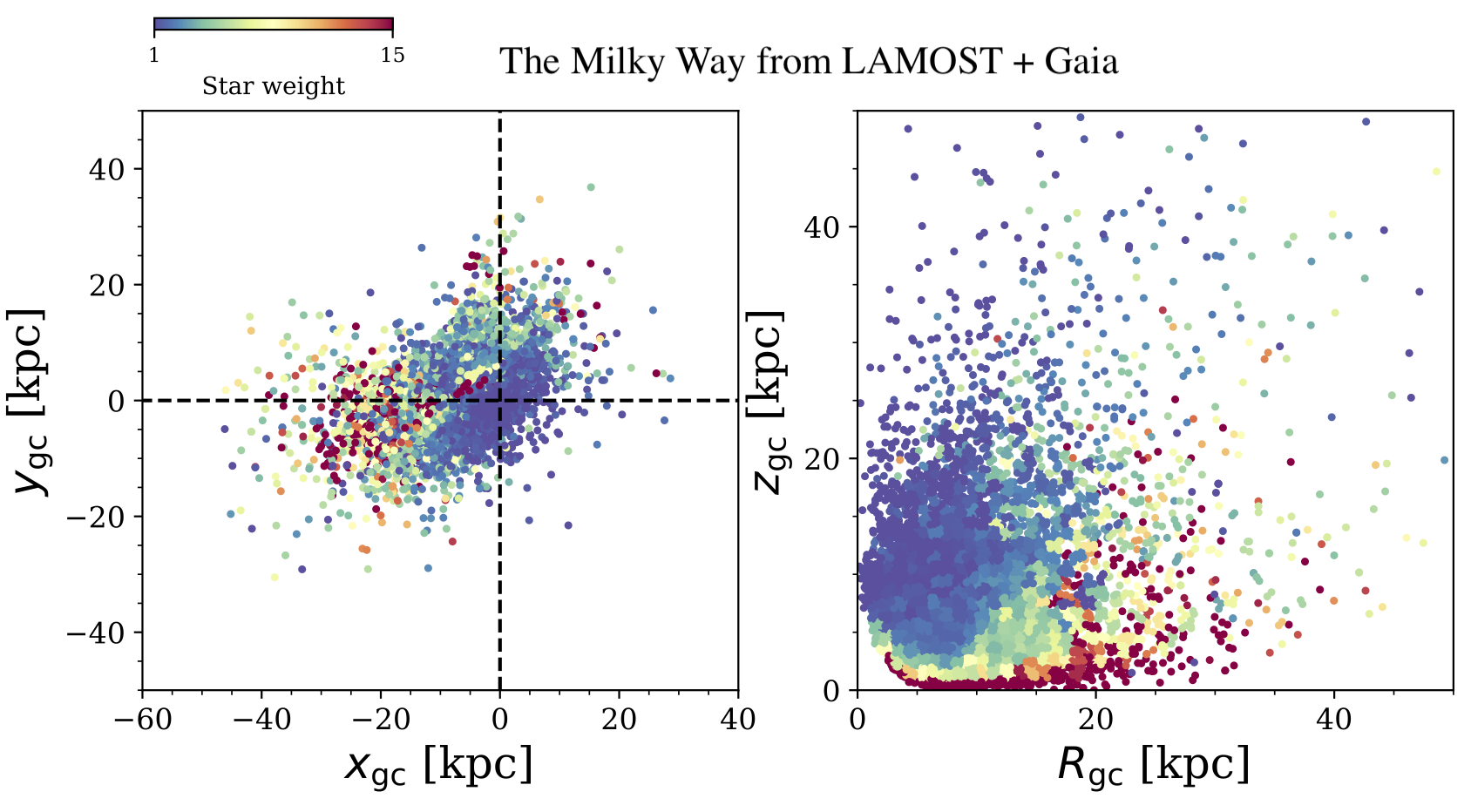}\centering\includegraphics[width=8.5cm]{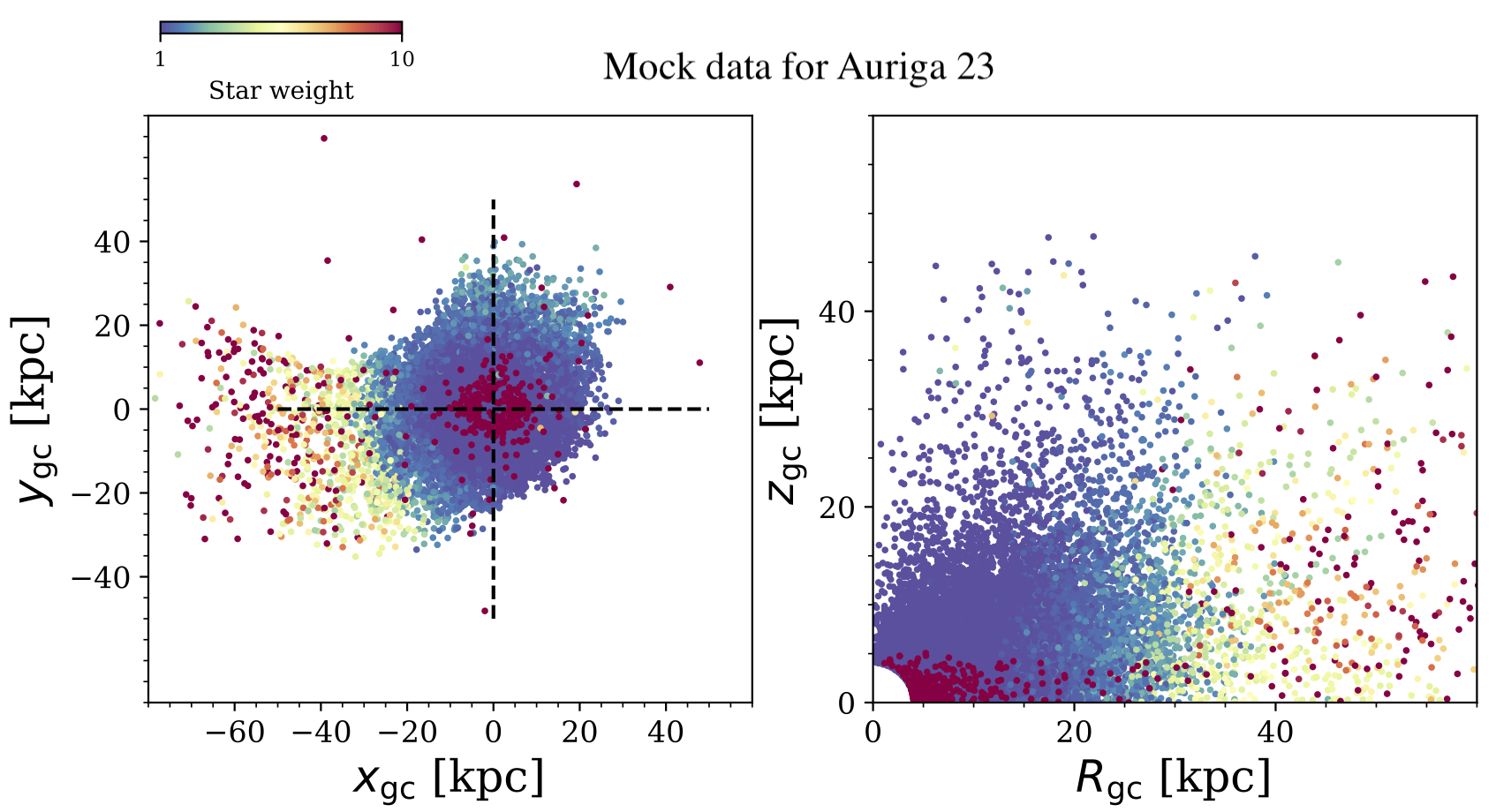}
\caption{{\bf Left:} MW observations from LAMOST + Gaia, and {\bf Right:} mock data created for Auriga 23. Each dot represents one star/particle colored by weight obtained from selection function correction. There are about 15,000 k-giants in the final sample of smooth halo of the Milky Way. The mock data of Auriga 23 includes about 20,000 particles, which generally match the spatial distribution of the MW sample.
}
\label{fig:sample}
\end{figure*}

\section{Simulations and mock data}
We first create mock data mimicking the LAMOST and Gaia observations from Auriga simulations. Then we process and refine the data to prepare them for dynamical modelling, following the same procedures applied to real observations.

\subsection{Auriga simulations}

The simulations used for our study are taken from the Auriga project \citep{Grand2017, Grand2019}, which is a suite of 40 cosmological magneto-hydrodynamical simulations for the formation of the Milky Way-mass haloes. These simulations were performed with the AREPO moving-mesh code \citep{Springel2010}, and follow many important galaxy formation processes such as star formation, a model for the ionising UV background radiation, a model for the multi-phase interstellar medium, mass loss and metal enrichment from stellar evolutionary processes, energetic supernovae and AGN feedback, and magnetic fields \citep{Pakmor2017}. We refer the reader to \citet{Grand2017} for more details. In this study, we select three galaxies from the Auriga simulation suite at a mass resolution of $\sim 5\times10^4\, M_{\odot}$ for baryons. The comoving gravitational softening length for the star particles and high-resolution dark matter particles is set to 500 $h^{-1}$ cpc. The physical gravitational softening length grows with the scale factor until a maximum physical softening length of 369 pc is reached. This corresponds to $z = 1$, after which the softening remains constant.

\subsubsection{Coordinates and measurement of true DM shape}
Initially, we find the principal axis of the stellar component. The galactocentric coordinate $\boldsymbol x_{\rm gc}\equiv \{x_{\rm gc},y_{\rm gc},z_{\rm gc}\}$ is established such that the \( z_{\rm gc} \)-axis is perpendicular to the disk, while the \( x_{\rm gc} \) and \( y_{\rm gc} \) axes are set somewhat arbitrarily within the plane of the disk, similar to the Milky Way's galactocentric coordinate. 

The intrinsic shapes of the dark matter (DM) halos in the Auriga simulation were determined by \citet{Prada2019}, using the axis ratio measured along the principal axis of the DM halo. In this analysis, we use Auriga 23, 5, and 12 to test our models. These three galaxies exhibit considerable variability in the intrinsic shapes of their DM halos, allowing us to assess our model's adaptability to various DM halo configurations.

The DM halo typically does not align precisely with the stars. The principal axes of the DM halo, denoted $\boldsymbol X\equiv \{X,Y,Z\}$, may be connected to the galactocentric coordinate through a rotation matrix, characterised by three Euler angles: $\alpha_q$, $\beta_q$, and $\gamma_q$:
\begin{equation}  \label{eq:halo_orientation}
\begin{aligned}
\boldsymbol X &= \mathsf R \boldsymbol x_{\rm gc}, \\
\mathsf R &\equiv \left(\!\! \begin{array}{ccc}  \phantom{-}
 c_\alpha c_\gamma - s_\alpha c_\beta s_\gamma\;\; & \phantom{-}
 s_\alpha c_\gamma + c_\alpha c_\beta s_\gamma\;\; &
 s_\beta  s_\gamma \\
-c_\alpha s_\gamma - s_\alpha c_\beta c_\gamma\;\; &
-s_\alpha s_\gamma + c_\alpha c_\beta c_\gamma\;\; &
 s_\beta  c_\gamma \\
 s_\alpha s_\beta &
-c_\alpha s_\beta &
 c_\beta 
\end{array} \!\!\right),
\end{aligned}
\end{equation}
where $c_\circ, s_\circ$ stands for $\cos\circ$, $\sin\circ$. The $Z$-axis is tilted by the angle $\beta_q$ relative to the $z_{\rm gc}$-axis. Meanwhile, the angles $\alpha_q$ ($\gamma_q$) define the rotation between the $x_{\rm gc}$ ($X$) axes and the intersection line of the $x_{\rm gc}y_{\rm gc}$ and $XY$ planes.

In practice, when DM distributions are derived from observational data, the orientation angles and intrinsic shapes of DM are highly degenerate. To simplify this, we set $Z=z_{\rm gc}$ by assuming $\beta_q = 0$\degree. The orientation of the DM halo in the $x_{\rm gc}y_{\rm gc}$ plane is left unrestricted. By projecting DM particles along the $Z$ axis, we determine the principal axis in the $XY$ plane using particles within $r<100$ kpc. The long axis in this plane is labelled as $X$ and the short axis as $Y$. We define the axis ratio as $p_{\rm DM} = Y/X$, and $q_{\rm DM}=Z/X$. In this definition, $p_{\rm DM}$ is restricted to be $<1$, while $q_{\rm DM}$ is allowed to be $>1$.

The intrinsic shape of Auriga halo 23, 5, and 12 measured in such a coordinate $XYZ$ is shown in Fig.~\ref{fig:vpq3_true}. The intrinsic shapes measured in our coordinate $XYZ$ are slightly different but close to the intrinsic shapes measured along the principal axes of the DM halo shown in \citet{Prada2019}.

\subsection{Mock data}
\label{SS:mock}
By cross-matching LAMOST with Gaia DR3, we have a sample of 619,284 k-giants with 3D velocity measured, the typical errors of radial velocity $v_r$, tangential velocity $v_{\phi}$, $v_{\theta}$ are 10, 30, 30 km/s, respectively. The typical distance error is about $15\%$. We exclude disk stars chemodynamically to avoid biasing the spatial distribution of the halo. We carefully exclude all substructures, including streams, overdensities, and GES using the friend-of-friend method \citep{Wang2022MNRAS.513.1958W, Yang2019ApJ...880...65Y}, and further perform $3\sigma$ clipping to remove outliers. In the end, we remain with 15,000 k-giants in the smooth halo with the spatial distribution shown in Fig.~\ref{fig:sample}. 
 
We take three galaxies from Auriga simulations, Auriga 23, 5, and 12. For each one, we take the following steps to create mock data to mimic LAMOST + Gaia observations for the MW halo. 

First, we treat each particle as a single star, transferring their positions and velocities into the ICRS coordinate. We adopt a relative error of $15\%$ for distance, an error of 30 km/s for the tangential velocity in the RA and DEC directions, and an error of 20 km/s for the LOS velocity. The positions and velocities of stars are then perturbed by random numbers, normally distributed with dispersions equal to the observational errors. 

Second, we perform spatial selection of the stars. 
We divide the particles into an original bulge with $r_{\rm gc}<4$ kpc, an original disk with $z_{\rm gc}<4$ kpc, $r_{\rm gc}>4$ kpc and $Z/Z_{\odot}>0.5$, and an original halo with the rest. We first randomly select 1/100 particles from the disk, 1/3 particles from the halo (1/4 for Auriga 12), and no particles from the bulge, then perform further spatial selection by only including stars in the northern hemisphere and within a spatial volume similar to the MW LAMOST + Gaia observations, as shown in Fig.\ref{fig:sample}. This results in $\sim 30000$ stars with the full 6D phase-space information measured. The sample size is much smaller than the original LAMOST + Gaia sample because we take a sparse sampling of the disk stars.

\begin{figure}
\centering\includegraphics[width=7cm]{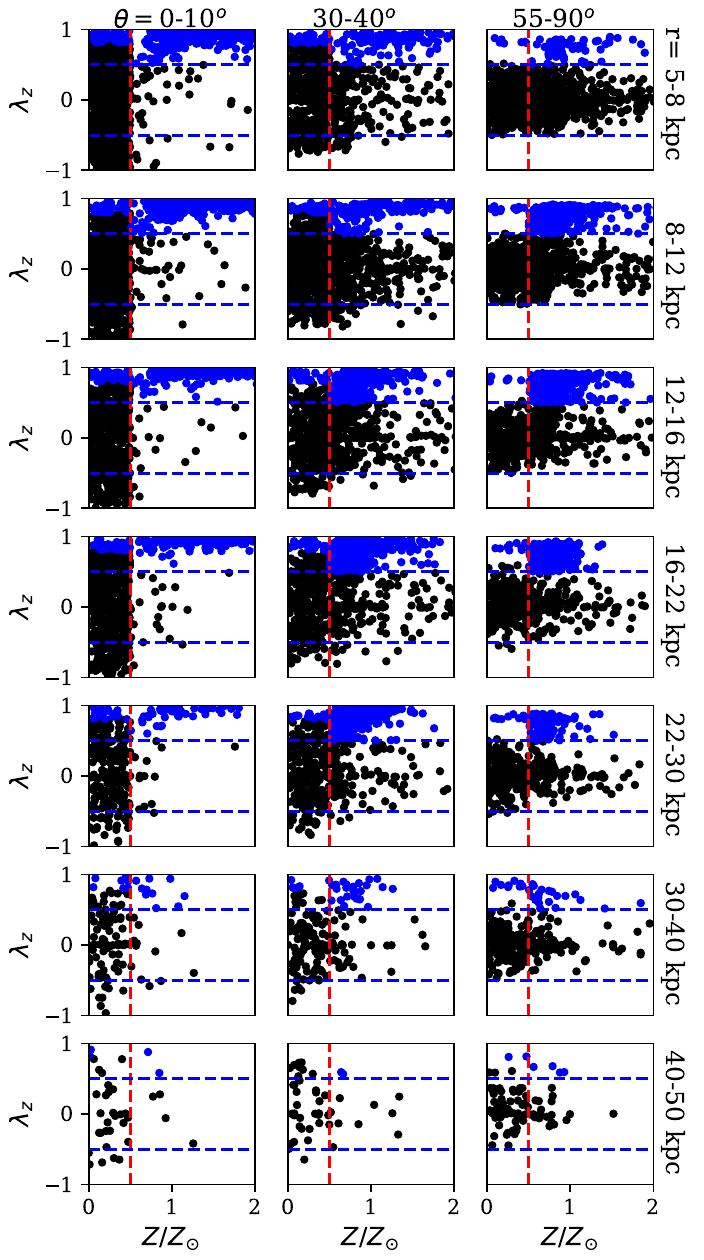}
\caption{Mock data for Auriga 23. The blue dots are disk stars that we excluded, i.e., stars with $\lambda_z>0.8$ or stars with $\lambda_z>0.5$ and $Z/Z_{\odot} > 0.5$. The black dots are taken as halo stars and kept in our sample. We have divided the data into $7 \times 6$ bins in $r_{\rm gc}$ versus $\theta$ with the intervals of $r_{\rm gc}=[5,8,12,16,22,30,40,50]$ kpc, $\theta$=[0,10,20,30,40,55,90\degree], here we only show three columns to illustrate the data.
}
\label{fig:zlz}
\end{figure}

\subsection{Data preparation and correction of selection function}
\label{SS:clean}
With mock data created in an observational manner, we convert the stars into the galactocentric coordinate and obtain the uncertainty of position and velocity in the new coordinate by a bootstrapping process.
We clean the data and correct the select function to make it ready for the model construction, in a way similar to how we deal with the real observations.

First, we use a position-velocity clustering estimator method to identify substructures; we select the smooth halo by removing all stars identified in substructures with a link length of 0.84 in the 6D distance \citep{Yang2019ApJ...880...65Y}. The stellar halos of Auriga 5 and 23 are rather smooth with no significant structures; we only perform this step on the mock data of Auriga 12. There are fewer substructures in the halo of Auriga simulations than in the real Milky Way galaxy due to the limited mass resolution in the simulation.

Second, we remove the stars from the disk. We integrate the stellar orbits in an initially guessed potential and calculate the circularity $\lambda_z$ of each orbit. $\lambda_z$ is defined as the angular momentum around the z axis normalised by the maximum allowed by circular orbits with the same binding energy:$\lambda_z = \overline{L_z}/(rV_c)$, where $\overline{L_z}=\overline{xv_y-yv_x}$, $r=\sqrt{\overline{x^2+y^2+z^2}}$, and $V_c = \sqrt{\overline{v_x^2 + v_y^2+v_z^2}}$, taking the average of the points along each orbit. A star in a near circular orbit has $\lambda_z \sim 1$, and on a radial orbit it has $\lambda_z\sim 0$. We remove the disk stars in a cut that combines circularity and metallicity; all the stars with $\lambda_z>0.8$ or $\lambda_z>0.5$ and $Z/Z_{\odot}>0.5$ are removed. As shown in Fig.~\ref{fig:zlz}, the remaining halo stars have almost a symmetric distribution in $\lambda_z$ centred at $\lambda_z=0$.

Third, we clean the data by $3\sigma$ clipping in the velocity distributions. Although we have removed the substructures identified by the clustering estimator method, there are still some particles with high tangential velocities which are unlikely to be part of a smooth halo in dynamical equilibrium. We also performed a $3\sigma$ clipping in $v_{\phi}$ versus $v_{\theta}$ to remove the outliers. 

Lastly, we correct for the selection function. For this proof-of-concept, we take a simple approach by correcting the selection function in the $R_{\rm gc}-z_{\rm gc}$ plane. We divide the $R_{\rm gc}-z_{\rm gc}$ plane into $50\times 50$ bins in $R_{\rm gc}=[0,50]$ kpc and $z_{\rm gc}=[0,50]$ kpc. We calculate the density of stars in each volume of $2\pi R_{\rm gc}dR_{\rm gc}dz_{\rm gc}$ constructed from our sample ($\rho_{\rm select}$) and that of the original halo before performing spatial selection ($\rho_{\rm original}$). Any star $i$ in the bin $j$ will have a weight of $w_i = \rho_{\rm original, j} / \rho_{\rm select, j}$. Such a correction of selection is less accurate at $z_{\rm gc}< 4$ kpc where disk stars are mixed in the original halo.

In the end, we have about 20,000 stars in the smooth halo; these stars are all located in the northern hemisphere and have the selection function corrected. The final sample of stars with their weights is shown in Fig.~\ref{fig:sample}.

The real selection function of LAMOST is complicated and is corrected by comparing with photometric data which are considered complete with the magnitude limit \citep{Liu2017RAA....17...96L, Yang2022AJ....164..241Y}, resulting in the correction of the selection function in the 3D space, more accurate than what we did here. It is beyond the scope of this paper to test the method of selection function correction. We just wanted the spatial sampling of our mock data to match some major properties of the real observations: (1) the data are not complete in different azimuthal angles, and we have less data in the direction of the Galactic Centre; (2) the observations at larger distances or with small polar angles are less complete; thus, we got higher weights for the stars at these positions.
As shown in Fig.~\ref{fig:sample}, the particle weight distribution in our mock data set generally matches the trend in the LAMOST + Gaia MW data.

\begin{figure*}
\centering\includegraphics[width=16cm]{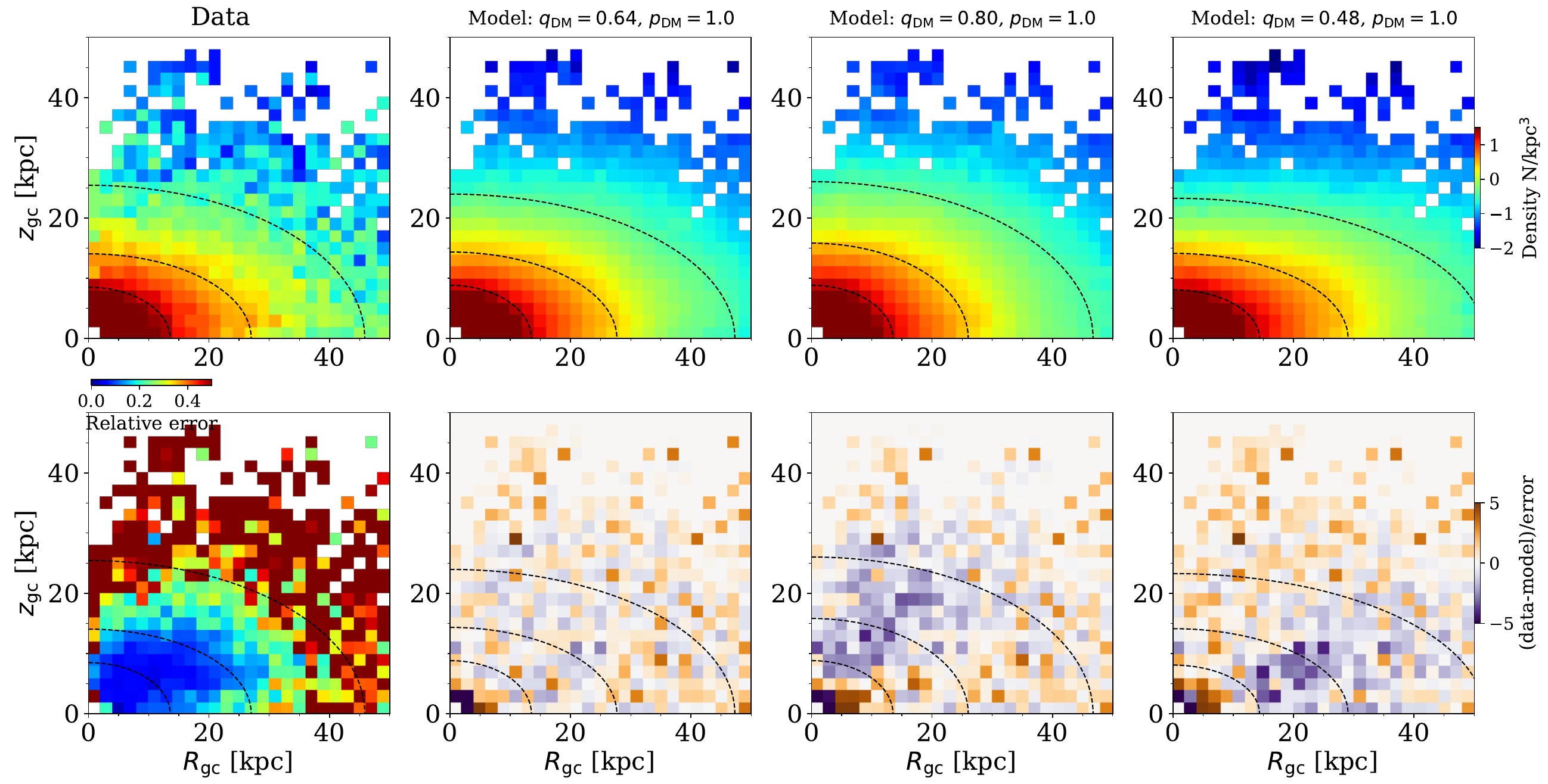}
\caption{Comparison of stellar density distribution from data and model for Auriga 23. The top panel from left to right are that constructed by observational data, and several models with different DM axis ratios of $q_{\rm DM}$ and $p_{\rm DM}$. The colors represent the number density as indicated by the colorbar, the dashed curves are iso-density contours. The second row are the uncertainty of the data, and model residuals. The data are well matched by the model with DM $q_{DM} = 0.64$ and $p_{DM} = 1$ which are well consistent with the ground truth. The stellar density distribution constructed by models are either too round or too flat if the shape of underlying DM changes. The stellar density distribution strongly constrains the 3D shape of the underlying DM halo.
}
\label{fig:SBmodel}
\end{figure*}

\begin{figure*}
\centering\includegraphics[width=5.5cm]{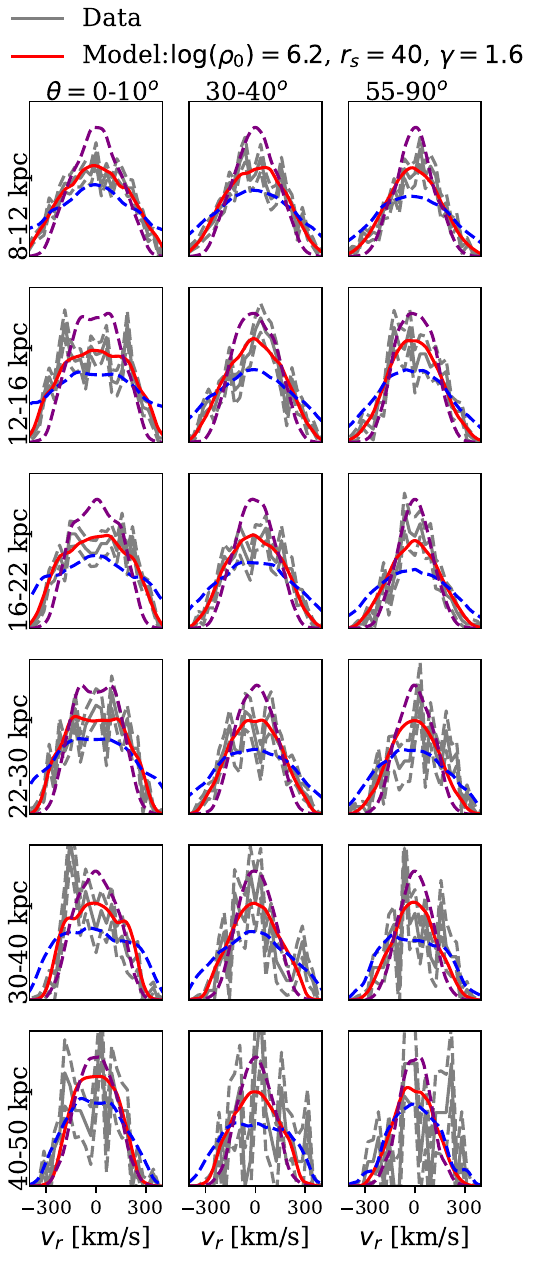}\centering\includegraphics[width=5.5cm]{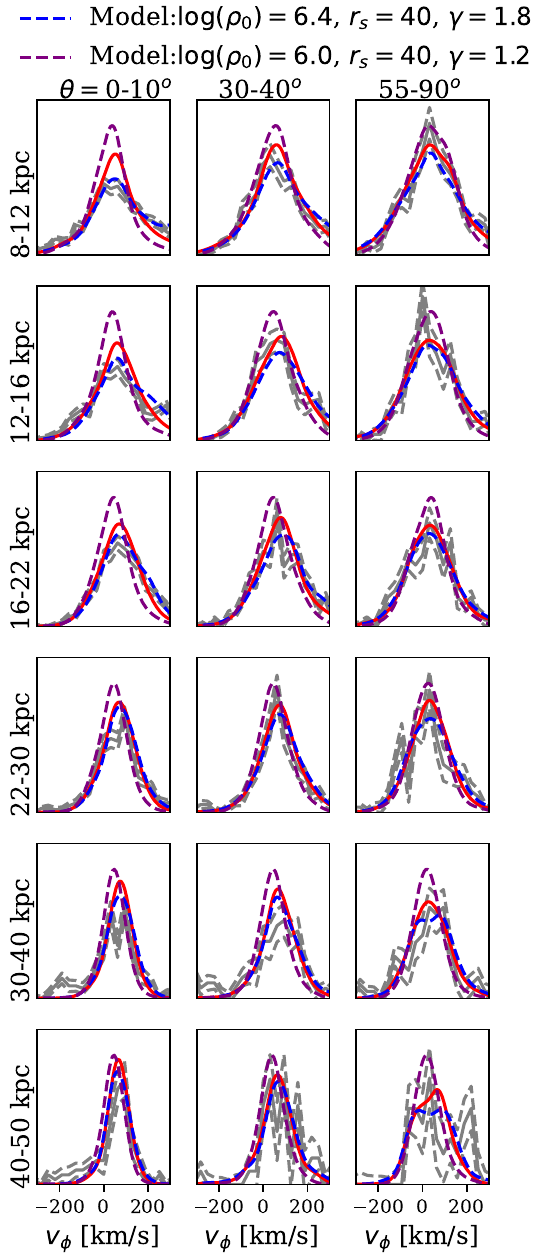}\centering\includegraphics[width=5.54cm]{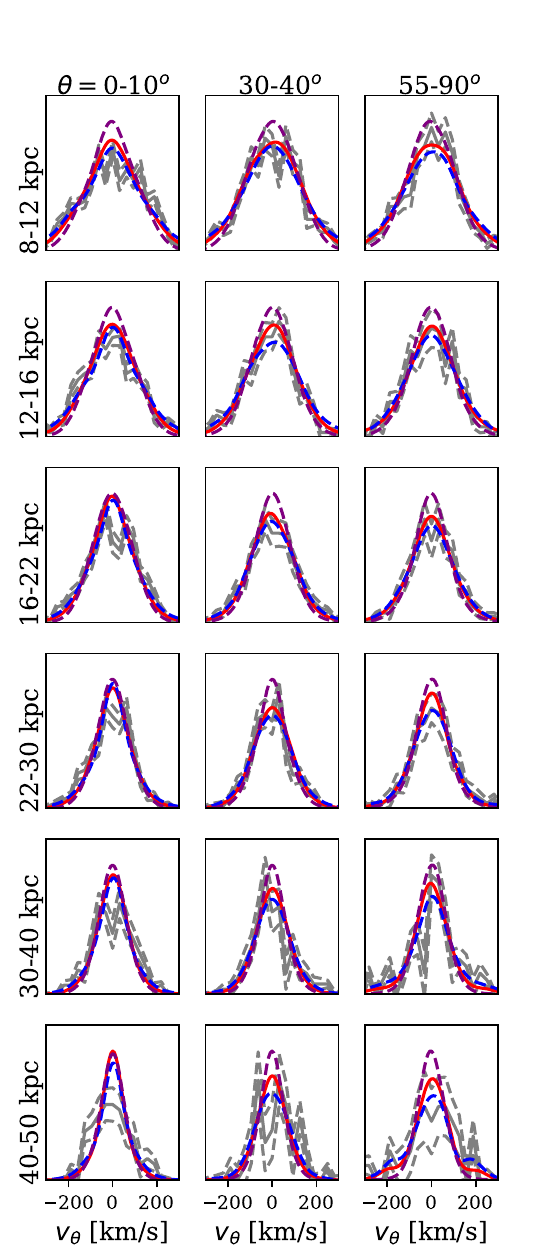}
\caption{Comparison of velocity distributions from data and model for Auriga 23. The velocity distribution $v_r$, $v_\phi$, and $v_{\theta}$ are calculated in $7\times 6$ bins in $r$ versus $\theta$, but we only show three columns here as labeled. In each panel, the gray solid and dashed curves are the velocity distribution and uncertainty of observational data. The red solid curves are from the best-fitting model, blue and magenta dashed curves are models with different radial distribution of the underlying DM mass. The velocity distributions, especially $v_r$, strongly constrain the underlying DM radial density distribution. The best-fitting model (red solid) well reproduces the velocity distributions in all three components: $v_r$, $v_{\phi}$, and $v_{\theta}$.
}
\label{fig:vhist}
\end{figure*}

\begin{figure*}
\centering\includegraphics[width=16cm]{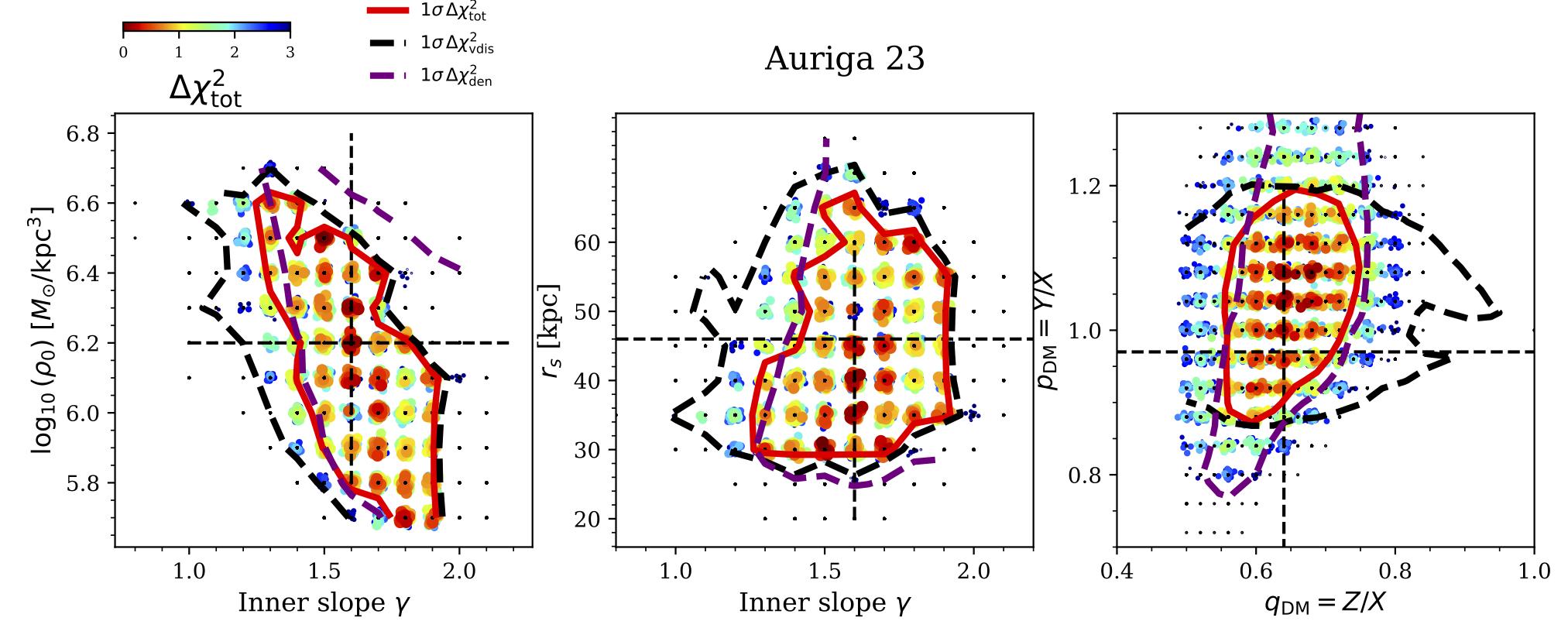}
\caption{Constraints on the parameters of the underlying DM distribution for Auriga 23. Each dot represents one model colored by the relative $\Delta \chi^2_{\rm tot}$ comparing to the least $\chi^2$ model. The parameters are sampled on grid with fixed step sizes, we slightly dithered the positions of the points to make them visible. The red solid, black, and purple dashed contours are the $1\sigma$ confidence level determined by $\Delta \chi^2_{\rm tot}$, $\Delta \chi^2_{\rm vdis}$, and $\Delta \chi^2_{\rm den}$, respectively.
The black horizontal and vertical dashed lines indicate the ground truth. The three parameters determining the radial distribution, $\rho_0$, $r_s$, $\gamma$, are degenerated. The shape of the DM halo $q_{\rm DM}$ and $p_{\rm DM}$ are well constrained and consistent with the ground truth.
}
\label{fig:grid}
\end{figure*}

\section{Model Construction}
We thus have a sample of stars in the halo, which can be considered complete in the northern hemisphere at $r_{\rm gc}\lesssim 50$ kpc after correction of the selection function. That is, we have the distribution function (DF) of halo stars in 6D phase space $f(\boldsymbol x,\boldsymbol v)$ numerically known from the observations. 
We take the assumption that the smooth halo is stationary; that the DF of stars integrated in the correct potential matches that directly observed instantaneously. And under this assumption, the DF constructed from the northern hemisphere should be representative of the whole system.

We construct a dynamical model, named Empirical Orbit-Superposition model, with the following steps: (1) we construct an adaptable gravitational potential for the MW with a few free parameters; (2) in a gravitational potential with given parameters, we calculate the stellar orbits by using stars' positions and velocities measured as initial conditions, and construct a model by superposing the stellar orbits using the weights of stars given by the selection function correction; (3) we extract stellar density distribution and velocity distributions from the orbit-superposition model, and evaluate the goodness of the model by calculating the likelihood ($\chi^2$) of each model against the data; (4) we explore the parameter grid of the gravitational potential and find the best-fitting models with the maximum likelihood (or least $\chi^2$). 

This method, in principle, is similar to the traditional Schwarzschild orbit-superposition method, but we have the true DF (orbit library and orbit weights) known from the observations. The only assumption we make in this model is that the system is stationary, which is the minimum assumption for all dynamical models. And we will confirm the feasibility of this assumption for the smooth halo with a good match between the model and the data, as we will show in Section~\ref{SS:bestmodel}. 

\subsection{Gravitational potential}
We construct a model of the gravitational potential by including a bulge, a disk, and a dark-matter halo.

We adopt a Sersic bulge:
\begin{equation}
\Sigma_{\rm bulge} = \Sigma_0 \exp(-b_n(R_{\rm gc}/a_{\rm bulge})^{1/n});
\end{equation}
and a MiyamotoNagai disk:
\begin{equation}
\Phi_{\rm disk} = - \frac{M_{\rm disk}}{\sqrt{R_{\rm gc}^2 + (a_{\rm disk}+\sqrt{z_{\rm gc}^2 + b_{\rm disk}^2})^2}},
\end{equation}
in which $R_{\rm gc} = \sqrt{x_{\rm gc}^2 + y_{\rm gc}^2}$. Our data in the halo do not have strong constraints on the bulge and disk. We fix the density distribution of the bulge and disks by the true values of the simulation \citep{Grand2017}. We have $M_{*,\rm bulge} = 3.2 \times 10^{10} M_{\odot}$, scale radius $a_{\rm bulge} = 1.71$ kpc, and s\'ersic index $n=1.44$, $M_{*,\rm disk} = 5.6 \times 10^{10} M_{\odot}$, scale radius $a_{\rm disk} = 4.9$ kpc, scale height $b_{\rm disk} = 0.25$ kpc for Auriga 23. The disk and bulge mass still have some uncertainty for the MW. We tried models with disk mass $20\%$ higher or lower for Auriga 23 and found no difference in our results. 
The true values parameterize the bulge and disk of Auriga 5 and Auriga 12 in \citet{Grand2017} are adopted.

We adopt a flexible triaxial generalised NFW model for the DM halo, allowing free orientations of the DM halo, and variable 3D shape as a function of radius generally following \citet{Vasiliev2021MNRAS.501.2279V}:
\begin{equation}
\rho_{\rm halo} = \rho_0 (\frac{\tilde{r}}{r_s})^{-\gamma} [1+(\frac{\tilde{r}}{r_s})^{\alpha}]^{\frac{\gamma-\beta}{\alpha}} \times \exp{-(\frac{\tilde{r}}{r_{\rm cut}})^{\xi}},
\end{equation}
in which $\tilde{r} =(p_{\rm DM}q_{\rm DM})^{1/3}\sqrt{X^2 + (Y/p_{\rm DM})^2 + (Z/q_{\rm DM})^2}$, $p_{\rm DM}$ and $q_{\rm DM}$ are the axis ratios of the DM halo. 

The coordinate $\boldsymbol X\equiv \{X,Y,Z\}$ is related to the galactocentric Cartesian coordinate $\boldsymbol x_{\rm gc}\equiv \{x_{\rm gc},y_{\rm gc},z_{\rm gc}\}$ by a rotation matrix parametrised by three Euler angles $\alpha_q$, $\beta_q$, $\gamma_q$ as shown in Section~\ref{SS:mock}. In principle, all three angles can be left free. Taking into account the degeneracy between the DM orientation and the 3D shape, we fix $\beta_q = 0$. The combination of $\alpha_q$ and $\gamma_q$ determines the DM halo orientation in the $XY$ plane; we fix $\gamma_q=0$ and allow $\alpha_q$ to be free to test the ability of constraining it. For default models, we also fix $\alpha_q = 0$.

When it is needed, we allow $p_{\rm DM}$ and $q_{\rm DM}$ to vary as a function of radius with:
\begin{eqnarray}
\label{eqn:pr}
p_{\rm DM}(r) = (p_{\rm in,DM} + p_{\rm out,DM} (\frac{r-10}{r_q})^2) / (1 + (\frac{r-10}{r_q} )^2)\\
p_{\rm DM}(r<10) = p_{\rm in,DM}\\
q_{\rm DM}(r) = (q_{\rm in,DM} + q_{\rm out,DM} (\frac{r-10}{r_q})^2) / (1 + (\frac{r-10}{r_q} )^2)\\
q_{\rm DM}(r<10) = q_{\rm in,DM}
\label{eqn:qr}
\end{eqnarray}
where $r = \sqrt{X^2 + Y^2 + Z^2}$, $p_{\rm in,DM}$, $p_{\rm out,DM}$, $q_{\rm in,DM}$, $q_{\rm out,DM}$, and the scale radius $r_q$ are allowed to be free parameters.

For the radial profile, we fixed $\alpha=1$, $\beta=3$, chose the outer cutoff radius $r_{\rm cut}=500$ kpc and the cutoff strength $\xi=5$. We start with models for all three mock galaxies with fixed DM orientation ($\alpha_q =0$, $\beta_q=0$ and $\gamma_q=0$) and constant $p_{\rm DM}$ and $q_{\rm DM}$. The five parameters in the DM halo: the scale density $\rho_0$ (or the mass $M_{\rm halo}$), the scale radius $r_s$, the inner density slope $\gamma$, $p_{\rm DM}$ and $q_{\rm DM}$ are left as free parameters. For Auriga 12, we further construct a model with different orientation angles $\alpha_q$ and a model that allows $p(r)$ and $q(r)$ to vary as a function of radius.

\subsection{Orbit integration}
Orbit integration is performed using the publicly released AGAMA package \citep{Vasiliev2019MNRAS.482.1525V}\footnote{https://github.com/GalacticDynamics-Oxford/Agama}. We have about 20,000 stars in the cleaned sample of the smooth halo, with their positions and velocities measured from observations, and the weight of each star obtained from the correction of selection. 
We start with the position and velocity of the stars from observations and integrate 10 orbital periods for each star. We withdraw 1000 particles from each orbit with equal time steps, and each particle is given the weight as the star that initialises the orbit. We superpose particles drawn from the orbits together, thus obtaining an orbit-superposed model, which numerically represents the DF of the stellar system. 

\subsection{Comparing model and data}
The DF of the orbit-superposed model, $f_{\rm model}(\boldsymbol x,\boldsymbol v | \rm potential)$, is based on data but incorporates the gravitational potential. If the system is in dynamical equilibrium, the DF of the model with the accurate gravitational potential should be statistically the same as that of the stars that initialised it $f_{\rm data}(\boldsymbol x,\boldsymbol v)$. We detail particular components of the DF: the spatial density distribution and velocity distributions at different positions. We evaluated the goodness of the model by comparing these with the data.

We calculate the stellar density distribution in the $R_{\rm gc}$-$z_{\rm gc}$ plane, by dividing it into $25\times25$ bins within 50 kpc. The density in each bin is calculated by $N/(2\pi R dR dz)$ in units of $N/{\rm kpc}^3$. We calculate the density directly for both the observed data $\rho_{\rm data}$ and the model $\rho_{\rm model}$. The uncertainty of the data $d\rho_{\rm data}$ is assessed by bootstrapping the position of stars with their uncertainties. The density distributions of the data and model are shown in Fig.~\ref{fig:SBmodel}; we calculate $\chi^2_{\rm den}$ by directly comparing the data and model:
\begin{equation}
\chi^2_{\rm den} = \sum (\rho_{\rm data} - \rho_{\rm model})^2/ d\rho_{\rm data}^2.
\end{equation}

To analyse velocity distributions, we categorise model particles into bins $N_{\rm bin} = 7 \times 6$ in $r$ and $\theta$. The divisions are $r = [5,8,12,16,22,30,40,50]$ kpc and $\theta$=[0,10,20,30,40,55,90\degree]. We then construct the velocity distributions ($v_r$, $v_{\phi}$, $v_{\theta}$) within each bin, as illustrated in Fig. 3. For each velocity component within bin $j$, we represent its distribution by a histogram using intervals $M$, labelled as ($v^k_{m,j}$, $h^k_{m,j}$), where $k$ corresponds to the velocities $v_r$, $v_{\phi}$, and $v_{\theta}$.

In bin $j$, we have $N_j$ stars, and for a star $i$ with velocity and measurement uncertainty ($v^k_i$, $\sigma^k_i$) within that bin, we determine its likelihood in comparison to the model for each of the three velocity components as follows:
\begin{equation}
P^k_{i, j} = \frac{1}{\sqrt{2\pi} \times \sigma^k_i} \times \frac{\sum_{m=1}^M h^k_{m,j} \times  \exp[- (v^k_i - v^k_{m,j})^2 / (2(\sigma^k_i)^2) ]} {\sum_{m=1}^M(h^k_{m,j})}
\end{equation}
The collective likelihood of all $N_j$ stars in this bin is given by $L_j^k = \sum_{i=1}^{N_{\rm j}} \log(P^k_{i,j})$, and the overall likelihood across all the bins is represented by $L^k = \sum_{j=7}^{N_{\rm bin}} L_j^k$. Due to a deficit of stars at $r<4$ kpc, which contributes to the distribution function (DF) of the inner bins, the DF for the innermost bins of our model remains incomplete. Therefore, we omit the first six bins within the range r=[5,8] in our likelihood calculations. The likelihood across the three velocity components is merged to derive the constraints from the velocity distributions.
\begin{equation}
 \chi^2_{\rm vdis} = -2(L^{v_r} + L^{v_{\phi}} + L^{v_{\theta}}).
\end{equation}
Finally, we combine the $\chi^2$ values from both the density distribution and velocity distributions to obtain
\begin{equation}
 \chi^2_{\rm tot} = -2(L^{v_r} + L^{v_{\phi}} + L^{v_{\theta}}) + \chi^2_{\rm den}.
\end{equation}

\begin{figure}
\centering\includegraphics[width=9cm]{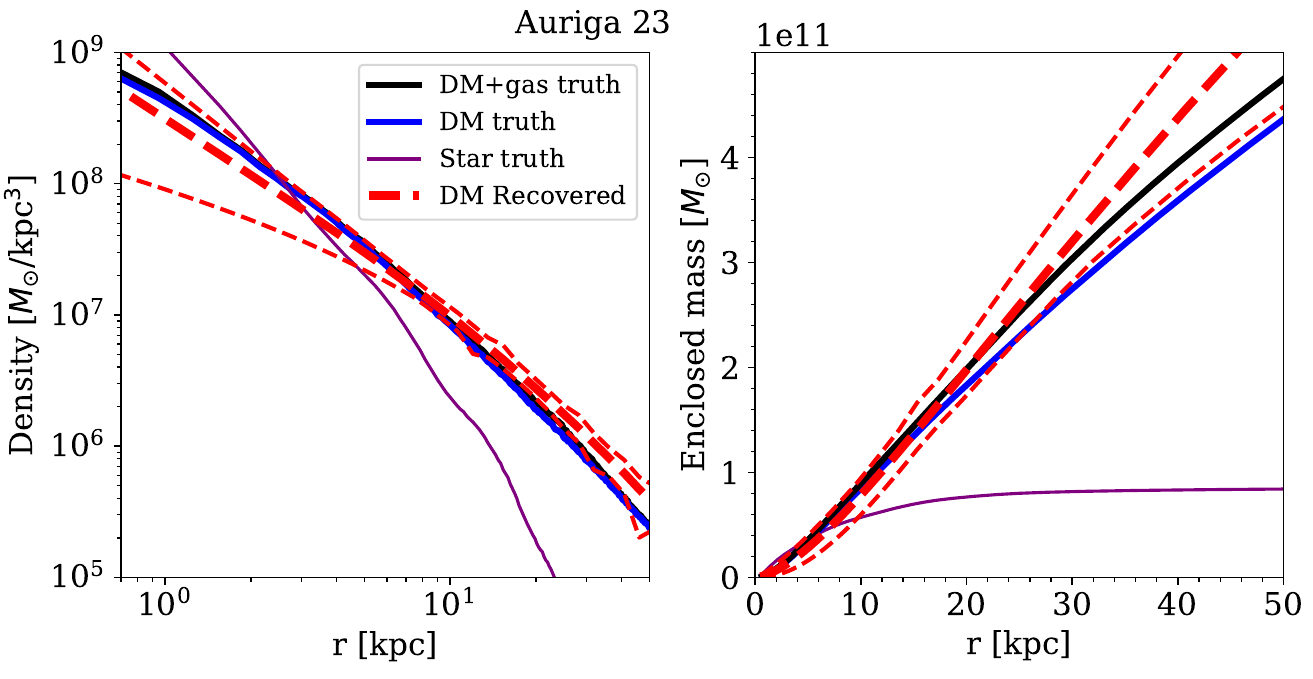}
\centering\includegraphics[width=9cm]{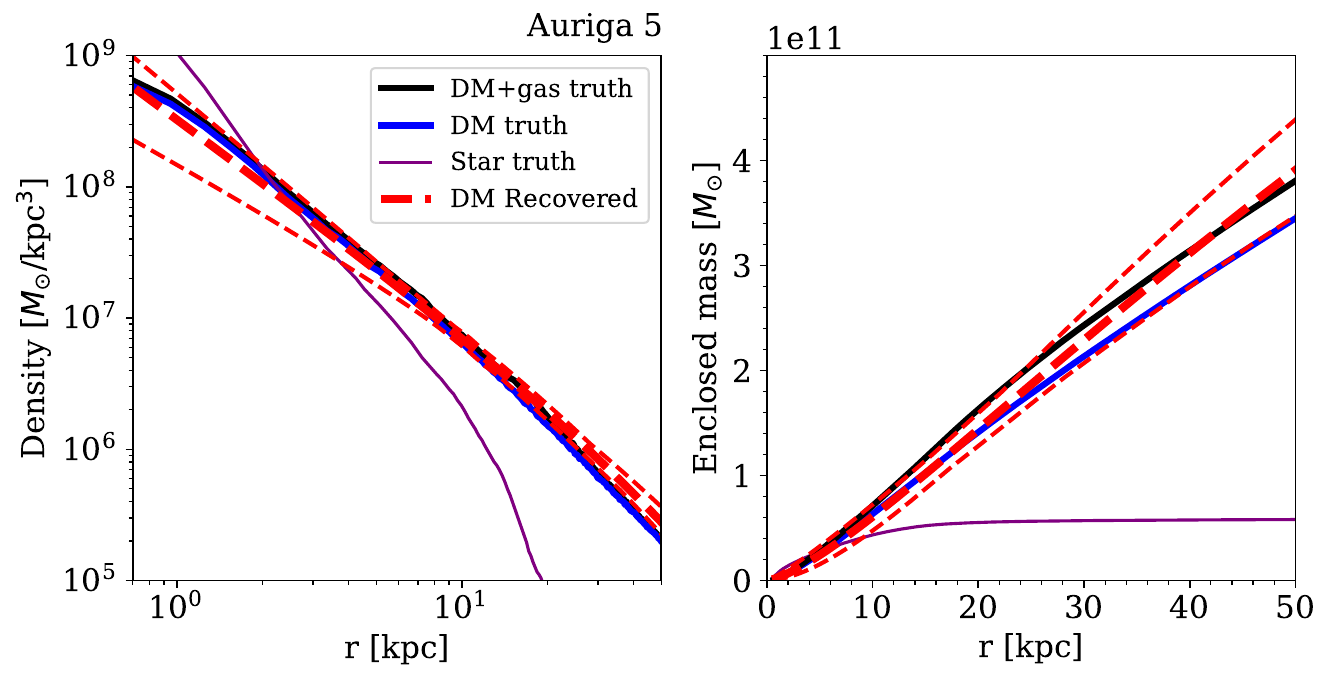}
\centering\includegraphics[width=9cm]{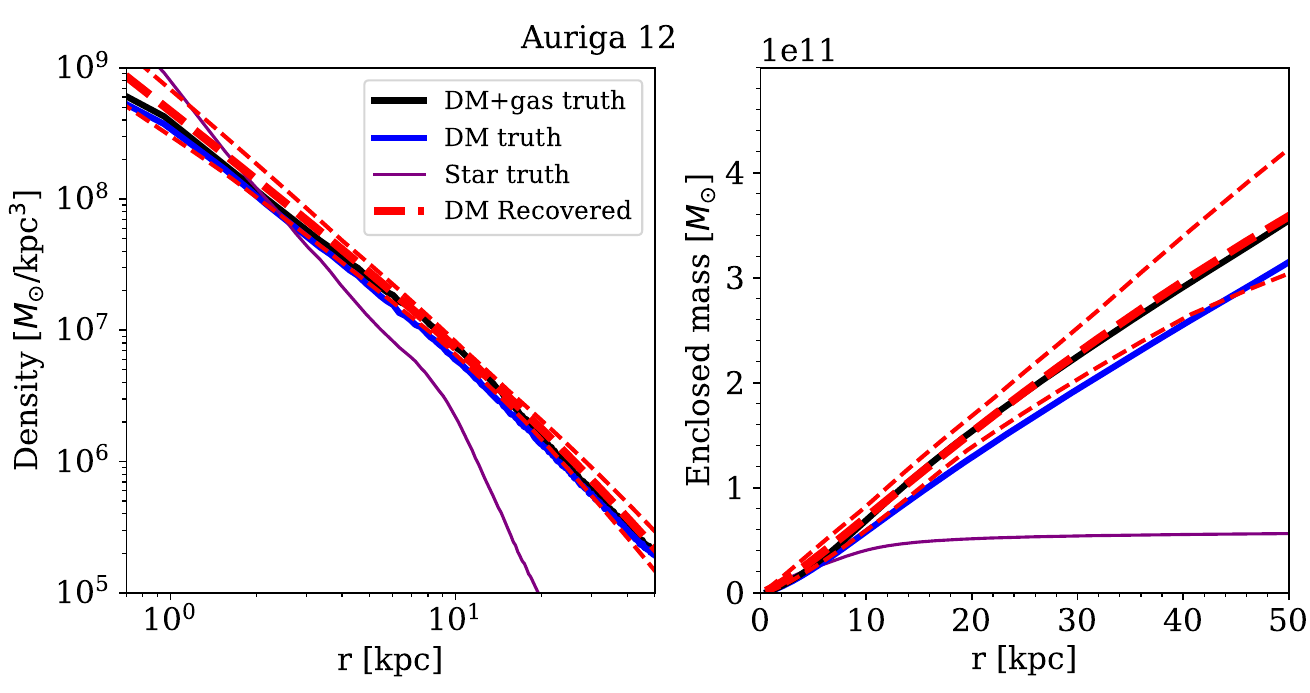}
\caption{The recovery of DM radial density distribution. The radial density profile and enclosed mass profile we obtained for Auriga 23, Auriga 5 and Auriga 12, which are well consistent with the ground truth within $1\sigma$ uncertainty. 
}
\label{fig:mass}
\end{figure}

\begin{figure*}
\centering\includegraphics[width=18cm]{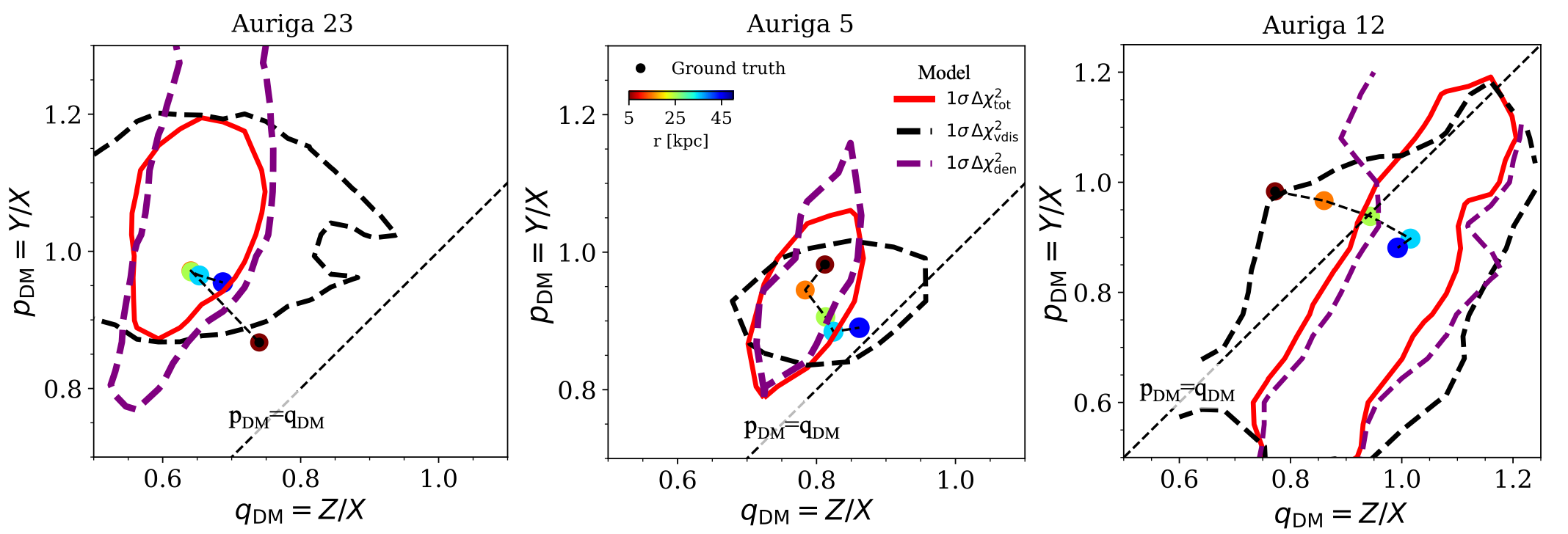}
\caption{The recovery of the DM intrinsic shapes by models with constant $q_{\rm DM}$ and $p_{\rm DM}$. The panels from left to right are Auriga 23, Auriga 5, and Auriga 12. In each panel, the red solid, black dashed, and purple dashed curves indicate the model $1\sigma$ confidence level determined by $\chi^2_{\rm tot}$, $\chi^2_{\rm vdis}$, and $\chi^2_{\rm den}$, respectively. The dots are the ground truth measured from the DM halo of the simulation at different radius, the colors represent the radii where $p_{\rm DM}$ and $q_{\rm DM}$ are measured as indicated by the colorbar. The DM halos of Auriga 23 and Auriga 5 are oblate with little variation from the inner to outer regions, which are well recovered by our model. The DM halo of Auriga 12 vary from oblate in the inner region to vertically aligned in the outer regions, the current models capture the shape of the DM at the outer regions.
}
\label{fig:pq3}
\end{figure*}

\begin{figure*}
\centering\includegraphics[width=18cm]{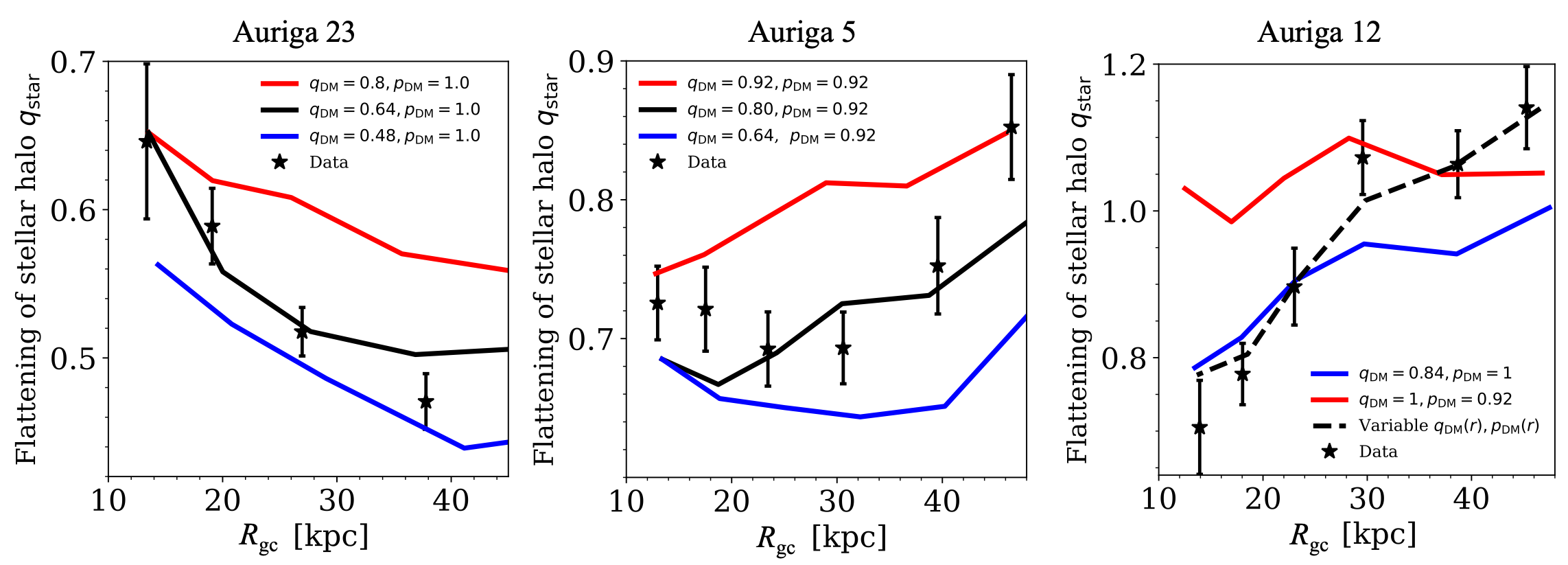}
\caption{Reproduce of the stellar halo flattening $q_{\rm star}$ as a function of radius $R_{\rm gc}$ for models with different underlying DM halos. The three panels from left to right are Auriga 23, 5, and 12. The DM halo shapes of Auriga 23 has little variation within our data coverage, its $q_{\rm star} (r)$ is well reproduced by a model with constant $p_{\rm DM}$ and $q_{\rm DM}$ close to the ground truth. Auriga 5 has mild variation, and Auriga 12 has relatively strong variation of the DM halo shape from $r\sim 10$ kpc to $r \sim 50$ kpc. Especially for Auriga 12, the stellar halo flattening $q_{\rm star} (r)$ at all regions can hardly be reproduced by a model with constant $p_{\rm DM}$ and $q_{\rm DM}$, in contrast, it can be reproduced well by a model with $p_{\rm DM}(r)$ and $q_{\rm DM}(r)$ vary with radius, close to its ground truth. The stellar halo flattening $q_{\rm star} (r)$ has strong constraints on the underlying DM shape.
}
\label{fig:pq3_qellp}
\end{figure*}

\subsection{Best-fitting models}
\label{SS:bestmodel}
We investigated the parameter space of $\rho_0$, $r_s$, $\gamma$, $p_{\rm DM}$, and $q_{\rm DM}$ within the gravitational potential. A parameter grid is established with intervals of 0.1, 5 kpc, 0.02, 0.04, and 0.04 for $\rho_0$, $r_s$, $\gamma$, $p_{\rm DM}$, and $q_{\rm DM}$, respectively. An iterative approach is used to find the optimal models. The process begins with an initial model, and, after developing initial models, iterative refinement follows. During each iteration, we identify the optimal models using the criterion $\chi^2 - \min (\chi^2) < \chi^2_s$, where $\chi^2_s = 100$. We then generate new models by walking two steps in every direction of the parameter grid of each optimal model. This approach guides the search towards the lower $\chi^2$ within the parameter grid and halts when the model with the minimum $\chi^2$ is identified. The relatively high value of $\chi^2_s$ is selected to ensure that all models within $1\sigma$ confidence level are considered before the end of the iteration. Ultimately, we determine the best-fitting models by achieving the minimum $\chi^2_{\rm tot}$.

We illustrate how the model works using Auriga 23 as an example. In Fig.~\ref{fig:SBmodel}, it is shown that the best-fitting model accurately reproduces the stellar density across the $R_{\rm gc}-z_{\rm gc}$ plane. Furthermore, it imposes strict constraints on the 3D shape of the underlying DM halo. The shape of the DM halo directly affects the direction of stellar acceleration and thus the shape of the stellar orbits. Variations in the DM halo shape can cause the constructed stellar halo to appear overly flat or excessively round. As shown in Fig.~\ref{fig:vhist}, the best-fitting model also reproduces the velocity distributions well in all three components: $v_r$, $v_{\phi}$, and $v_{\theta}$. These velocity distributions provide strong constraints on the radial density distribution of the DM halo which directly affects the strength of the acceleration, especially in the radial direction. The different radial profiles of the DM halo lead to notable discrepancies in the velocities between the observational data and the model, particularly in $v_r$. Of course, both the stellar density and velocity distributions are determined by the overall 3D DM distribution, which will be better constrained by combining both. 

Our best-fitting model generally reproduces both the density and velocity distributions at various locations in the $R_{\rm gc}-z_{\rm gc}$ plane, thus validating our assumption of an overall stationary system with the agreement between the model and the data. However, these stellar system are never in perfect equilibrium. Although we have tried our best to clean the substructures, there will inevitably be some residual substructures or non-equilibrium features left in the data. For example, in Fig.~\ref{fig:vhist}, there are some peaks/bumps in the velocity distributions that are not matched by the model, some of them can indicate residual substructures. The $v_{\phi}$ distributions in the bin at $8<r<12$ kpc close to the disk plane are not perfectly matched by the model either, which can indicate some non-equilibrium features. 


In Fig.~\ref{fig:grid}, we illustrate the parameter space explored for Auriga 23. The coloured points denote models within a confidence level $3\sigma$, colored by the normalised $\Delta \chi^2_{\rm tot} = (\chi^2_{\rm tot} - \min (\chi^2_{\rm tot}) )/ \Delta \chi^2_{1\sigma} $, where $\Delta \chi^2_{1\sigma} $ is the confidence level $1\sigma$ determined by bootstrapping \footnote{ We take the gravitational potential of the best-fitting model, and perturb the position and velocity of observed stars with their uncertainty for 100 trials. In each trial, we recompute the model and evaluate the resulting $\chi^2_{\rm tot}$ (including $\chi^2_{\rm den}$ and $\chi^2_{\rm vdis}$). The standard deviation of these 100 values $\chi^2$ is taken as the confidence level $1\sigma$ of the model $\Delta \chi^2_{1\sigma} $.}. As shown in Fig.~\ref{fig:grid}, the combination of density distribution and velocity distribution strongly constrains the 3D shape of DM $p_{\rm DM}$ and $q_{\rm DM}$. Due to the still limited data coverage, there are still strong degeneracies between the three parameters ($\rho_0$, $r_s$, and $\gamma$) determining the radial profile.  Nevertheless, the radial density profile within the data coverage is well constrained, as will be detailed in the next section.

\section{Results}
We have applied the model to mock data created from Auriga 23, Auriga 5 and Auriga 12. In this section, we will show our recovery of the DM radial density profile and 3D shape for these three galaxies. 
\subsection{Recovery of DM density profile}
We demonstrate how effectively our models recover the DM radial density profile and the enclosed DM mass in Fig.~\ref{fig:mass}. Despite significant degeneracy between $\rho_0$, $r_s$, and $\gamma$, our models accurately capture the radial density profile within the observational range of 4-50 kpc. The uncertainty is larger at $r<8$ kpc due to the absence of data. 

The enclosed DM mass profile is also well recovered, with a relative uncertainty of about $10\%$. It should be noted that the DM mass we obtained in the outer regions is $5\%-10\%$ higher than the true values but aligns well with the enclosed mass that combines DM and gas. The CircumGalactic Medium (CGM) in the halo of the simulation mirrors the DM's density distribution. The mass of CGM is compensated by an overestimate of the DM mass without a separate gas contribution in our model of the gravitational potential.

\begin{figure*}
\centering\includegraphics[width=8.7cm]{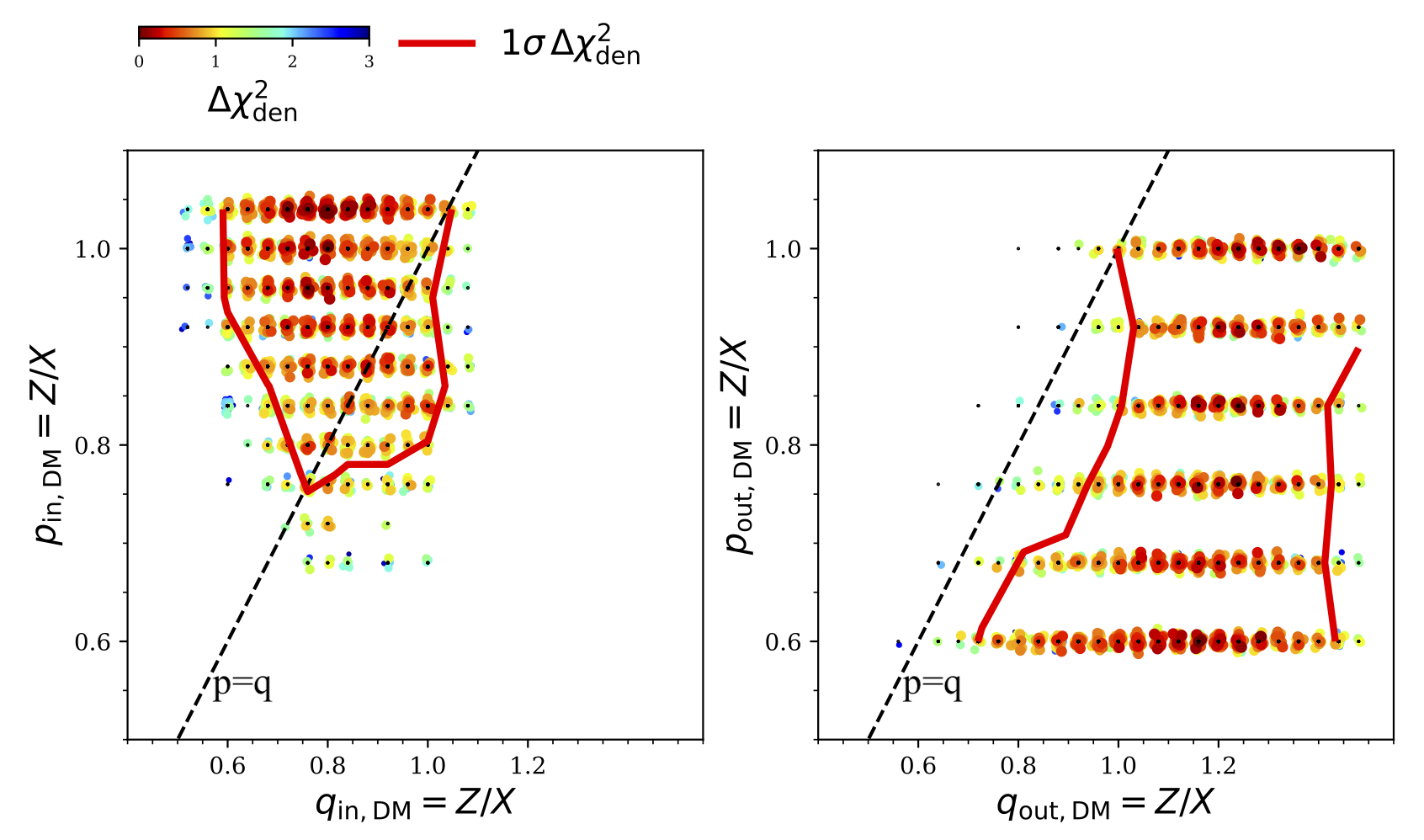}\centering\includegraphics[width=9cm]{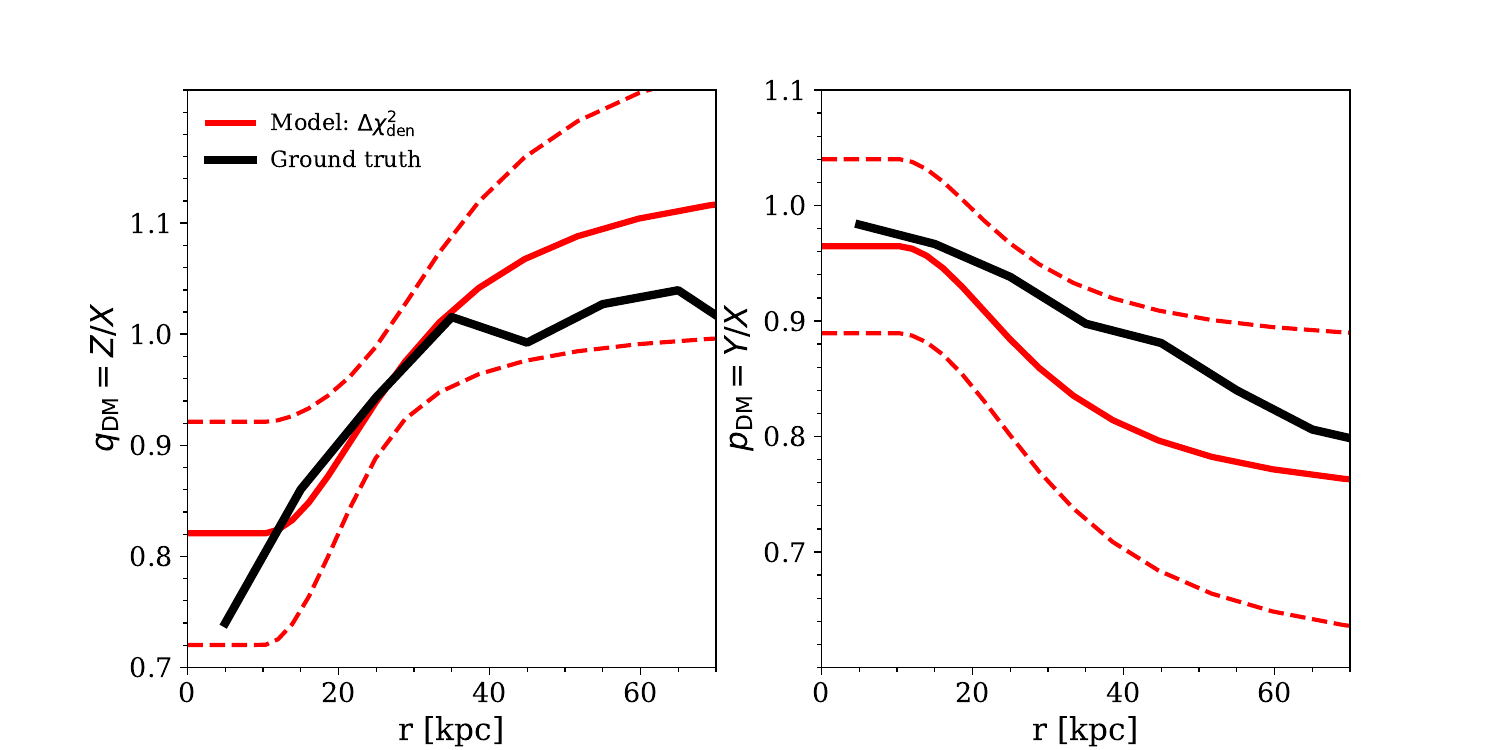}
\caption{{\bf Left:} the constraints on $q_{\rm in, DM}$, $p_{\rm in, DM}$, $q_{\rm out, DM}$ and $p_{\rm out, DM}$ for the model allowing DM axis ratios vary as a function of radius for Auriga 12. {\bf Right:} 
Comparison of $p_{\rm DM}(r)$ and $q_{\rm DM}(r)$ profile obtained by the model (red) and the ground truth (black). The red solid and dashed curves represent the mean and $1\sigma$ uncertainty of the results obtained by the model.
}
\label{fig:vpq}
\end{figure*}

\begin{figure}
\centering\includegraphics[width=9cm]{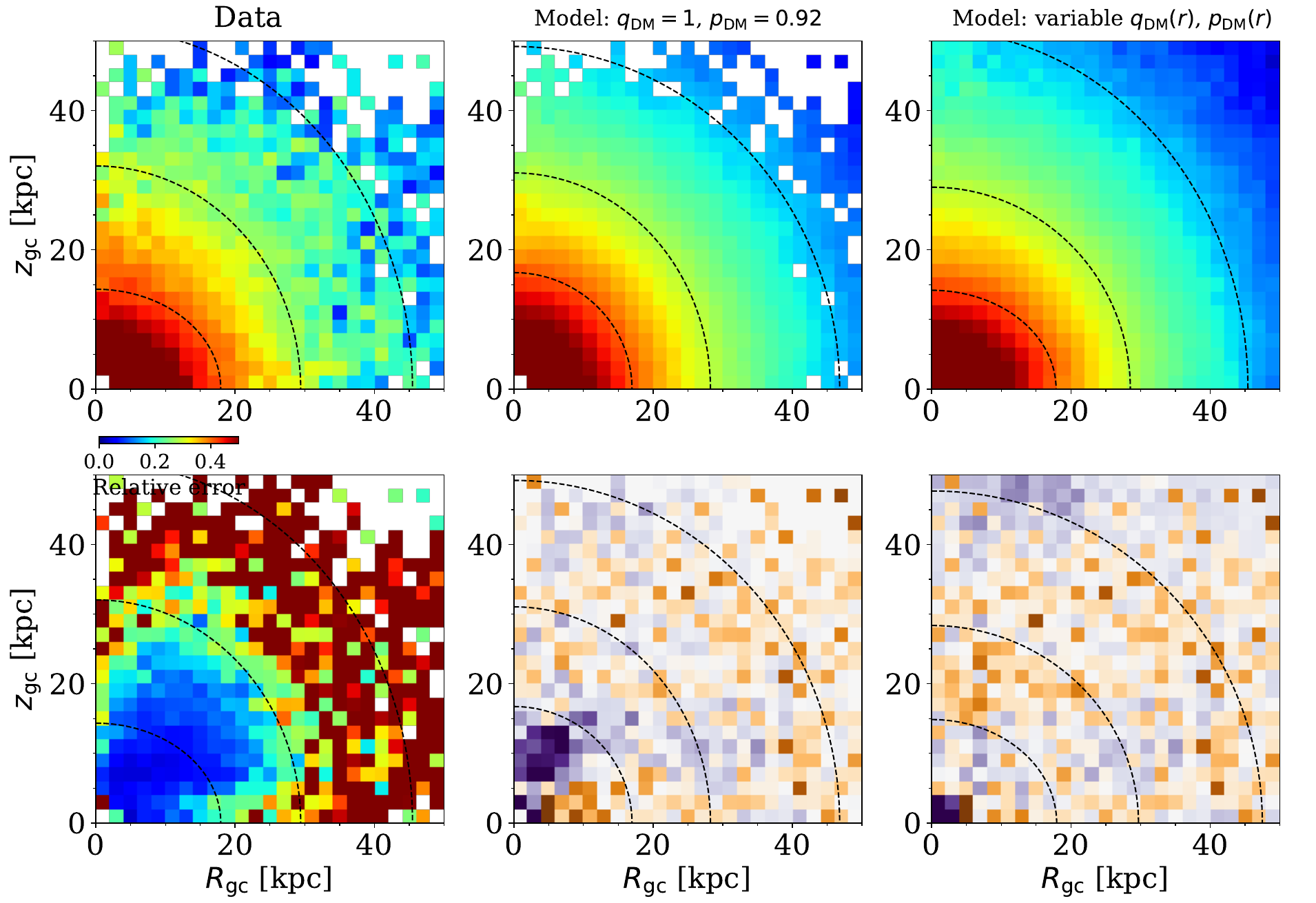}
\caption{Comparison of the stellar density distribution constructed from data and different models for Auriga 12. The columns from left to right are the data, best-fitting model with constant $p_{\rm DM}$ and $q_{\rm DM}$, best-fitting model with $p_{\rm DM}(r)$ and $q_{\rm DM}(r)$ varying as a function of radius $r$. The model with constant $p_{\rm DM}$ and $q_{\rm DM}$ can not match the density distribution all all radial regions, there are significant residuals in the inner 20 kpc. The consistency between data and model improves with the model allowing variable $p_{\rm DM}(r)$ and $q_{\rm DM}(r)$.
}
\label{fig:Au12SBmodel}
\end{figure}

\subsection{Recovery of the 3D shape of DM}
\subsubsection{Models with constant $q_{\rm DM}$ and $p_{\rm DM}$}

The three selected galaxies exhibit varied DM halo structures, as illustrated in Fig.~\ref{fig:vpq3_true}. The DM halo of Auriga 23 is predominantly oblate, showing minimal change from the inner to the outer regions. The DM halo of Auriga 5 is more triaxial, displaying slight variation. On the other hand, the DM halo of Auriga 12 undergoes notable variations, transitioning from oblate within the inner 20 kpc to a vertical orientation at $r>20$ kpc. 

Initially, we developed models with constant $q_{\rm DM}$ and $p_{\rm DM}$ for all three galaxies. As depicted in Fig.~\ref{fig:pq3}, our model successfully recovered both $p_{\rm DM}$ and $q_{\rm DM}$ for Auriga 23 and Auriga 5, with relatively small uncertainties. For Auriga 12, the model captured the outer configuration of the DM halo, and there is a significant degeneracy between $p_{\rm DM}$ and $q_{\rm DM}$ when the halo is vertically aligned, characterised by $p_{\rm DM}<q_{\rm DM}$.

The stellar density distribution in the $R_{\rm gc}-z_{\rm gc}$ plane, especially the flattening of the stellar halo, is crucial to constrain the DM shape. We measure the stellar halo flattening, $q_{\rm star}$, across various radii using isodensity contours and compared these from the data and several models in Fig.~\ref{fig:pq3_qellp}. The stellar halo of Auriga 23 is notably flat, becoming increasingly so from the centre to the outskirts, and our best-fitting model successfully mimicked $q_{\rm star}$ at all radii. The stellar halo of Auriga 5 appears rounder, with a tendency to become rounder moving outward; our model approximates the radial variation $q_{\rm star}(r)$, albeit with moderate inner region discrepancies. The stellar halo of Auriga 12 is the most spherical, notably rounding from inner to outer areas, with $q_{\rm star}(r)$ challenging to replicate by a model with a constant DM shape. The failure to reproduce the tendency of $q_{\rm star}(r)$ suggests a variation of DM shapes from the inner to outer regions.

\subsubsection{Models allowing DM shape to vary from inner to outer regions}
We create models with variables $p_{\rm DM}(r)$ and $q_{\rm DM}(r)$ as specified in Eq.~\ref{eqn:pr} to Eq.~\ref{eqn:qr} for Auriga 12. We fix the transition radius $r_q = 20$ kpc, while the remaining four parameters, $q_{\rm in, DM}$, $q_{\rm out, DM}$, $p_{\rm in, DM}$, and $p_{\rm out, DM}$, are treated as free parameters. The parameters determining the radial DM profile do not degenerate significantly with the 3D shape. To optimise the efficiency of the model, we fix them at $\log(\rho_0 [M_{\odot}/{\rm kpc}^3])=6.2$, $r_s=40$ kpc, and $\gamma=1.6$. 

The parameter grid we explored for this model is shown in Fig.\ref{fig:vpq}. 
Despite significant uncertainties, a twisted DM halo that is oblate ($p_{\rm in, DM} > q_{\rm in, DM}$ and $p_{\rm in, DM} \sim 1$) in the inner regions and vertically aligned in the outer regions ($p_{\rm out, DM} < q_{\rm out, DM}$ and $q_{\rm out, DM} \sim 1$) is strongly favoured. 

We derive $p_{\rm DM}(r)$ and $q_{\rm DM}(r)$ from our models within the $1\sigma$ confidence level, and compare those with the true profiles derived from the simulation. Our model generally recovers the true tendency of the DM shape variation with radius, as shown in the right panels of Fig.\ref{fig:vpq}.

Such models allowing variables $p_{\rm DM}(r)$ and $q_{\rm DM}(r)$ are indeed able to reproduce the stellar halo density distribution. We take one of the best fitting models with $p_{\rm in} = 1.04$, $p_{\rm out} = 0.84$, $q_{\rm in} = 0.6$ and $q_{\rm out} =1.32$, which reproduces the surface density distribution well from the inner to outer regions, with the residuals significantly reduced compared to the model with constant $p_{\rm DM}$ and $q_{\rm DM}$ as shown in Fig.~\ref{fig:Au12SBmodel}. The tendency of $q_{\rm star}(r)$ derived from this model matches the data well (the dashed black curve in Fig.~\ref{fig:pq3_qellp}).

\section{Discussion}
\label{s:dis}
We have fixed the DM orientation with the tilt angle $\beta_q=0$. In reality, the DM halo might not be perfectly aligned with the stellar disk. The three galaxies we tested have DM tilt angles of $\beta_q \sim 10-15$\degree. The axis ratio $q$ recovered from our model is close to the true axis ratio measured along the fixed $z_{\rm gc}$ axis, not the intrinsic $q$ measured along the principal axes of the DM halo, although the latter two are close to each other when $\beta_q$ is small. For galaxies with a larger tilt angle $\beta_q$, we expect the model to still recover $q$ measured along the fixed $z_{\rm gc}$ in a similar way; in this case, it may be significantly different from the intrinsic $q$ measured along the DM principal axes.

The orientation of the DM halo in the $x_{\rm gc}y_{\rm gc}$ plane remains largely undetermined. Ideally, if the DM orientation $\alpha_q$ could be constrained, we will converge to two solutions, one with $p<1$ and the correct $\alpha_q$ and one with $p' = 1/p$ when $\alpha_q$ is misaligned with 90\degree. In our exploration of the Auriga 12 model that allows a free orientation ($\alpha_q$) of the DM halo in the $x_{\rm gc}y_{\rm gc}$ plane, we failed to find significant constraints on $\alpha_q$. In particular, the stellar density distributions created by various $\alpha_q$ values of the DM halo appeared similar, resulting in no constraints on $\alpha_q$ from $\chi^2_{\rm den}$, and the same is true for the radial velocity distribution $v_r$. The velocities $v_{\phi}$ and $v_{\theta}$ are somewhat influenced by the orientation of the DM $\alpha_q$, but this effect occurs in a somewhat unpredictable manner. 

This arises because we averaged out information along the azimuthal angle when correcting the selection function and comparing the data with the model. We have to do so because observational data at various azimuthal angles remain incomplete, particularly with a lack of observations directed towards the Galactic Centre. 
Consequently, this process results in a significant level of uncertainty for $p_{\rm DM} = Y/X$, especially when $p_{\rm DM}<q_{\rm DM}$, as demonstrated by Auriga 12. Despite this, the relative positioning of $p_{\rm DM}$ and $q_{\rm DM}$ is effectively constrained. In our models for Auriga 12, $p_{\rm DM} < q_{\rm DM}$ is strongly preferred in the outer regions, in contrast to Auriga 23 and 5, where $p_{\rm DM}>q_{\rm DM}$ are strongly preferred.

Stellar halo of the three galaxies we tested is generally in dynamical equilibrium, but inevitably with some residual substructures and non-equilibrium features; they did not introduce obvious systematic bias according to our test results.
The LMC, which fell in recently, may have perturbed the Milky Way halo and will affect the results of dynamical modelling. However, LMC mainly induces a bulk motion in the outskirts of the halo, while only mildly affects the inner halo, causing a velocity shift in $V_z$ with $\sim 10$ km/s at $r<50$ kpc \citep{Erkal2020MNRAS.498.5574E}. Among the Auriga simulations, Auriga 25 has a recent minor merger with mass and orbits similar to LMC, but
its perturbation to the disk and inner halo is significantly
stronger than the LMC did to the Milky Way. We apply the method to Auriga 25 and show that even with some obvious non-equilibrium in the data, the DM distribution can still be roughly recovered but with larger statistical uncertainty (see the Appendix).

\section{Summary}
\label{s:conclusion}

We present an empirical triaxial orbit-superposition model tailored for the Milky Way halo. In this method, we take the minimum assumption of dynamical
models that the stellar halo is stationary, meaning that the distribution function (DF) in the 6D phase space, $f(\boldsymbol x,\boldsymbol v)$, does not change when the stars orbit in the correct gravitational potential. The feasibility of this assumption can be gauged by the goodness of the match of the data and the model in the end. 

The gravitational potential of the model is a combination of disk, bulge, and DM halo with adaptable parameters. Our approach employs a highly flexible triaxial NFW halo, which allows variations in the DM axis ratios ($p_{\rm DM} = Y/X$, $q_{\rm DM} = Z/X$) with radius. Within each specified gravitational potential, we take the 6D phase-space information of stars observed as initial conditions, integrate their orbits in the gravitational potential, and form a model by superposing the orbits together. The orbits are weighted by the weights assigned to the stars when correcting the observation selection function. We evaluated the goodness of the model by comparing the DF of the model with that directly constructed from the data, expressed in terms of density distribution and velocity distributions ($v_{r}$, $v_{\phi}$, $v_{\theta}$) across the $R_{\rm gc}-z_{\rm gc}$ plane.

This methodology is validated using mock observational data of the Milky Way halo from LAMOST + Gaia with three Auriga galaxies, halo 23, 5, 12 with different intrinsic shapes of the DM halo. The mock data are created with similar spatial selection and observational errors as in LAMOST + Gaia. We clean the data by subtracting the disk, substructures, etc., in exactly the same way as we deal with the real observational data. 

We create models with constant DM axis ratios to all three mock galaxies and extra models allowing $p_{\rm DM}(r)$, $q_{\rm DM}(r)$ varying with radius for Auriga 12. We find that:

\begin{itemize}
    \item The density distribution and velocity distributions of the stellar halo provide consistent constraints on the DM distribution. Specifically, the stellar density distribution has stronger constraints on the DM shape, while velocity distributions better restrict the DM radial density profile.  
    \item The DM radial density profile and the enclosed DM mass of the three galaxies within the data coverage are well recovered, with a relative uncertainty of about $10\%$. Although we still have relatively large uncertainty in each of the three parameters that determine the radial profile ($\rho_0$, $r_s$, and $\gamma$) with limited data coverage.
    \item Auriga 23 and Auriga 5 have oblate to triaxial DM halo that vary only mildly with radius. The DM axis ratios ($p_{\rm DM}$, $q_{\rm DM}$) are well recovered by the models with constant $p_{\rm DM}$ and $q_{\rm DM}$, and the best-fitting models reproduce the stellar halo density distribution reasonably well. 
    \item The DM halo of Auriga 12 is twisted, oblate ($p_{\rm DM}>q_{\rm DM}$) in the inner 20 kpc, and vertically aligned ($p_{\rm DM}<q_{\rm DM}$) in the outer regions. Our model with constant $p_{\rm DM}$ and $q_{\rm DM}$ recovers those in the outer regions, and it does not reproduce the stellar halo density distribution in the inner regions well. We recover $p_{\rm DM}(r)$, $q_{\rm DM}(r)$ of Auriga 12 as a function of radius generally well with a model that allows them to vary, which also better reproduces the stellar halo density distribution from inner to outer regions.
\end{itemize}

\begin{acknowledgement}
We thank David Hogg, Jianhui Lian, Zhaozhou Li for useful discussions. LZ acknowledges the support from the CAS Project for Young Scientists in Basic Research under grant No. YSBR-062 and the National Key R\&D Program of China No. 2022YFF0503403. X.X.X. acknowledges the support from the National Key Research and Development Program of China No. 2024YFA1611902, CAS Project for Young Scientists in Basic Research grant No. YSBR-062 and YSBR-092, and the science research grants from the China Manned Space Project with NO. CMS-CSST-2025-A11.
\end{acknowledgement}

\bibliographystyle{aa}  
\bibliography{ms_relic} 

\begin{thebibliography}{50}
\expandafter\ifx\csname natexlab\endcsname\relax\def\natexlab#1{#1}\fi

\bibitem[{{Bird} {et~al.}(2022){Bird}, {Xue}, {Liu}, {Flynn}, {Shen}, {Wang}, {Yang}, {Zhai}, {Zhu}, {Zhao}, \& {Tian}}]{Bird2022MNRAS.516..731B}
{Bird}, S.~A., {Xue}, X.-X., {Liu}, C., {et~al.} 2022, \mnras, 516, 731

\bibitem[{{Bovy} {et~al.}(2016){Bovy}, {Bahmanyar}, {Fritz}, \& {Kallivayalil}}]{Bovy2016ApJ...833...31B}
{Bovy}, J., {Bahmanyar}, A., {Fritz}, T.~K., \& {Kallivayalil}, N. 2016, \apj, 833, 31

\bibitem[{{Bovy} {et~al.}(2018){Bovy}, {Kawata}, \& {Hunt}}]{Bovy2018MNRAS.473.2288B}
{Bovy}, J., {Kawata}, D., \& {Hunt}, J. A.~S. 2018, \mnras, 473, 2288

\bibitem[{{Bowden} {et~al.}(2015){Bowden}, {Belokurov}, \& {Evans}}]{Bowden2015MNRAS.449.1391B}
{Bowden}, A., {Belokurov}, V., \& {Evans}, N.~W. 2015, \mnras, 449, 1391

\bibitem[{{Erkal} {et~al.}(2020){Erkal}, {Belokurov}, \& {Parkin}}]{Erkal2020MNRAS.498.5574E}
{Erkal}, D., {Belokurov}, V.~A., \& {Parkin}, D.~L. 2020, \mnras, 498, 5574

\bibitem[{{Grand} {et~al.}(2017){Grand}, {Gomez}, {Marinacci}, {Pakmor}, {Springel}, {Campbell}, {Frenk}, {Jenkins}, \& {White}}]{Grand2017}
{Grand}, R.~J.~J., {Gomez}, F.~A., {Marinacci}, F., {et~al.} 2017, \mnras, 467, 179

\bibitem[{{Grand} {et~al.}(2019){Grand}, {van de Voort}, {Zjupa}, {Fragkoudi}, {G{\'o}mez}, {Kauffmann}, {Marinacci}, {Pakmor}, {Springel}, \& {White}}]{Grand2019}
{Grand}, R. J.~J., {van de Voort}, F., {Zjupa}, J., {et~al.} 2019, \mnras, 490, 4786

\bibitem[{{Green} {et~al.}(2023){Green}, {Ting}, \& {Kamdar}}]{Green2023ApJ...942...26G}
{Green}, G.~M., {Ting}, Y.-S., \& {Kamdar}, H. 2023, \apj, 942, 26

\bibitem[{{Han} {et~al.}(2016){Han}, {Wang}, {Cole}, \& {Frenk}}]{Han2016MNRAS.456.1017H}
{Han}, J., {Wang}, W., {Cole}, S., \& {Frenk}, C.~S. 2016, \mnras, 456, 1017

\bibitem[{{Hattori} {et~al.}(2021){Hattori}, {Valluri}, \& {Vasiliev}}]{Hattori2021MNRAS.508.5468H}
{Hattori}, K., {Valluri}, M., \& {Vasiliev}, E. 2021, \mnras, 508, 5468

\bibitem[{{Helmi}(2004)}]{Helmi2004ApJ...610L..97H}
{Helmi}, A. 2004, \apjl, 610, L97

\bibitem[{{Ibata} {et~al.}(2001){Ibata}, {Lewis}, {Irwin}, {Totten}, \& {Quinn}}]{Ibata2001ApJ...551..294I}
{Ibata}, R., {Lewis}, G.~F., {Irwin}, M., {Totten}, E., \& {Quinn}, T. 2001, \apj, 551, 294

\bibitem[{{Johnston} {et~al.}(2005){Johnston}, {Law}, \& {Majewski}}]{Johnston2005ApJ...619..800J}
{Johnston}, K.~V., {Law}, D.~R., \& {Majewski}, S.~R. 2005, \apj, 619, 800

\bibitem[{{Khoperskov} {et~al.}(2024{\natexlab{a}}){Khoperskov}, {Di Matteo}, {Steinmetz}, {Ratcliffe}, {van de Ven}, {Boin}, {Haywood}, {Kacharov}, {Minchev}, {Krajnovic}, {Valentini}, \& {de Jong}}]{Khobulge2024arXiv241118182K}
{Khoperskov}, S., {Di Matteo}, P., {Steinmetz}, M., {et~al.} 2024{\natexlab{a}}, arXiv e-prints, arXiv:2411.18182

\bibitem[{{Khoperskov} {et~al.}(2024{\natexlab{b}}){Khoperskov}, {Steinmetz}, {Haywood}, {van de Ven}, {Krajnovic}, {Ratcliffe}, {Minchev}, {Di Matteo}, {Kacharov}, {Marques}, {Valentini}, \& {de Jong}}]{Khodisk2024arXiv241116866K}
{Khoperskov}, S., {Steinmetz}, M., {Haywood}, M., {et~al.} 2024{\natexlab{b}}, arXiv e-prints, arXiv:2411.16866

\bibitem[{{Khoperskov} {et~al.}(2024{\natexlab{c}}){Khoperskov}, {van de Ven}, {Steinmetz}, {Ratcliffe}, {Minchev}, {Krajnovic}, {Haywood}, {Di Matteo}, {Kacharov}, {Marques}, {Valentini}, \& {de Jong}}]{kho2024arXiv241115062K}
{Khoperskov}, S., {van de Ven}, G., {Steinmetz}, M., {et~al.} 2024{\natexlab{c}}, arXiv e-prints, arXiv:2411.15062

\bibitem[{{Koposov} {et~al.}(2010){Koposov}, {Rix}, \& {Hogg}}]{Koposov2010ApJ...712..260K}
{Koposov}, S.~E., {Rix}, H.-W., \& {Hogg}, D.~W. 2010, \apj, 712, 260

\bibitem[{{K{\"u}pper} {et~al.}(2015){K{\"u}pper}, {Balbinot}, {Bonaca}, {Johnston}, {Hogg}, {Kroupa}, \& {Santiago}}]{kupper2015ApJ...803...80K}
{K{\"u}pper}, A. H.~W., {Balbinot}, E., {Bonaca}, A., {et~al.} 2015, \apj, 803, 80

\bibitem[{{Law} \& {Majewski}(2010)}]{Law2010ApJ...714..229L}
{Law}, D.~R. \& {Majewski}, S.~R. 2010, \apj, 714, 229

\bibitem[{{Li} \& {Binney}(2022)}]{Li2022MNRAS.510.4706L}
{Li}, C. \& {Binney}, J. 2022, \mnras, 510, 4706

\bibitem[{{Li} {et~al.}(2024){Li}, {Han}, {Wang}, {Qian}, {Li}, {Jing}, \& {Li}}]{Li2024arXiv240811414L}
{Li}, Z., {Han}, J., {Wang}, W., {et~al.} 2024, arXiv e-prints, arXiv:2408.11414

\bibitem[{{Liu} {et~al.}(2017){Liu}, {Xu}, {Wan}, {Wang}, {Carlin}, {Deng}, {Newberg}, {Cao}, {Hou}, {Wang}, \& {Zhang}}]{Liu2017RAA....17...96L}
{Liu}, C., {Xu}, Y., {Wan}, J.-C., {et~al.} 2017, Research in Astronomy and Astrophysics, 17, 096

\bibitem[{{Loebman} {et~al.}(2014){Loebman}, {Ivezi{\'c}}, {Quinn}, {Bovy}, {Christensen}, {Juri{\'c}}, {Ro{\v{s}}kar}, {Brooks}, \& {Governato}}]{Loebman2014ApJ...794..151L}
{Loebman}, S.~R., {Ivezi{\'c}}, {\v{Z}}., {Quinn}, T.~R., {et~al.} 2014, \apj, 794, 151

\bibitem[{{Long} {et~al.}(2013){Long}, {Mao}, {Shen}, \& {Wang}}]{Long2013MNRAS.428.3478L}
{Long}, R.~J., {Mao}, S., {Shen}, J., \& {Wang}, Y. 2013, \mnras, 428, 3478

\bibitem[{{Malhan} \& {Ibata}(2019)}]{Malhan2019MNRAS.486.2995M}
{Malhan}, K. \& {Ibata}, R.~A. 2019, \mnras, 486, 2995

\bibitem[{{Nitschai} {et~al.}(2020){Nitschai}, {Cappellari}, \& {Neumayer}}]{Nitschai2020MNRAS.494.6001N}
{Nitschai}, M.~S., {Cappellari}, M., \& {Neumayer}, N. 2020, \mnras, 494, 6001

\bibitem[{{Pakmor} {et~al.}(2017){Pakmor}, {G{\'o}mez}, {Grand }, {Marinacci}, {Simpson}, {Springel}, {Campbell}, {Frenk}, {Guillet}, {Pfrommer}, \& {White}}]{Pakmor2017}
{Pakmor}, R., {G{\'o}mez}, F.~A., {Grand }, R. J.~J., {et~al.} 2017, \mnras, 469, 3185

\bibitem[{{Palau} \& {Miralda-Escud{\'e}}(2019)}]{Palau2019MNRAS.488.1535P}
{Palau}, C.~G. \& {Miralda-Escud{\'e}}, J. 2019, \mnras, 488, 1535

\bibitem[{{Palau} \& {Miralda-Escud{\'e}}(2023)}]{Palau2023MNRAS.524.2124P}
{Palau}, C.~G. \& {Miralda-Escud{\'e}}, J. 2023, \mnras, 524, 2124

\bibitem[{{Panithanpaisal} {et~al.}(2022){Panithanpaisal}, {Sanderson}, {Arora}, {Cunningham}, \& {Baptista}}]{Pani2022arXiv221014983P}
{Panithanpaisal}, N., {Sanderson}, R.~E., {Arora}, A., {Cunningham}, E.~C., \& {Baptista}, J. 2022, arXiv e-prints, arXiv:2210.14983

\bibitem[{{Portail} {et~al.}(2017){Portail}, {Gerhard}, {Wegg}, \& {Ness}}]{Portail2017MNRAS.465.1621P}
{Portail}, M., {Gerhard}, O., {Wegg}, C., \& {Ness}, M. 2017, \mnras, 465, 1621

\bibitem[{{Posti} \& {Helmi}(2019)}]{Posti2019A&A...621A..56P}
{Posti}, L. \& {Helmi}, A. 2019, \aap, 621, A56

\bibitem[{{Prada} {et~al.}(2019){Prada}, {Forero-Romero}, {Grand}, {Pakmor}, \& {Springel}}]{Prada2019}
{Prada}, J., {Forero-Romero}, J.~E., {Grand}, R. J.~J., {Pakmor}, R., \& {Springel}, V. 2019, \mnras, 490, 4877

\bibitem[{{Springel}(2010)}]{Springel2010}
{Springel}, V. 2010, \mnras, 401, 791

\bibitem[{{Tahmasebzadeh} {et~al.}(2024){Tahmasebzadeh}, {Zhu}, {Shen}, {Gadotti}, {Valluri}, {Thater}, {van de Ven}, {Jin}, {Gerhard}, {Erwin}, {Jethwa}, {Zocchi}, {Lilley}, {Fragkoudi}, {de Lorenzo-C{\'a}ceres}, {M{\'e}ndez-Abreu}, {Neumann}, \& {Guo}}]{Tah2024MNRAS.534..861T}
{Tahmasebzadeh}, B., {Zhu}, L., {Shen}, J., {et~al.} 2024, \mnras, 534, 861

\bibitem[{{Thater} {et~al.}(2022){Thater}, {Jethwa}, {Tahmasebzadeh}, {Zhu}, {den Brok}, {Santucci}, {Ding}, {Poci}, {Lilley}, {Tim de Zeeuw}, {Zocchi}, {Maindl}, {Rigamonti}, {Yang}, {Fahrion}, \& {van de Ven}}]{Thater2022A&A...667A..51T}
{Thater}, S., {Jethwa}, P., {Tahmasebzadeh}, B., {et~al.} 2022, \aap, 667, A51

\bibitem[{{van den Bosch} {et~al.}(2008){van den Bosch}, {van de Ven}, {Verolme}, {Cappellari}, \& {de Zeeuw}}]{vdB2008}
{van den Bosch}, R.~C.~E., {van de Ven}, G., {Verolme}, E.~K., {Cappellari}, M., \& {de Zeeuw}, P.~T. 2008, \mnras, 385, 647

\bibitem[{{Vasiliev}(2019)}]{Vasiliev2019MNRAS.482.1525V}
{Vasiliev}, E. 2019, \mnras, 482, 1525

\bibitem[{{Vasiliev} \& {Athanassoula}(2015)}]{Vasiliev.2015}
{Vasiliev}, E. \& {Athanassoula}, E. 2015, \mnras, 450, 2842

\bibitem[{{Vasiliev} {et~al.}(2021){Vasiliev}, {Belokurov}, \& {Erkal}}]{Vasiliev2021MNRAS.501.2279V}
{Vasiliev}, E., {Belokurov}, V., \& {Erkal}, D. 2021, \mnras, 501, 2279

\bibitem[{{Vera-Ciro} \& {Helmi}(2013)}]{Vera2013ApJ...773L...4V}
{Vera-Ciro}, C. \& {Helmi}, A. 2013, \apjl, 773, L4

\bibitem[{{Wang} {et~al.}(2022){Wang}, {Zhang}, {Xue}, {Huang}, {Liu}, {Zhang}, \& {Yang}}]{Wang2022MNRAS.513.1958W}
{Wang}, F., {Zhang}, H.~W., {Xue}, X.~X., {et~al.} 2022, \mnras, 513, 1958

\bibitem[{{Wang} {et~al.}(2020){Wang}, {Han}, {Cautun}, {Li}, \& {Ishigaki}}]{Wang2020SCPMA..6309801W}
{Wang}, W., {Han}, J., {Cautun}, M., {Li}, Z., \& {Ishigaki}, M.~N. 2020, Science China Physics, Mechanics, and Astronomy, 63, 109801

\bibitem[{{Wegg} {et~al.}(2019){Wegg}, {Gerhard}, \& {Bieth}}]{Wegg2019MNRAS.485.3296W}
{Wegg}, C., {Gerhard}, O., \& {Bieth}, M. 2019, \mnras, 485, 3296

\bibitem[{{Woudenberg} \& {Helmi}(2024)}]{Woud2024A&A...691A.277W}
{Woudenberg}, H.~C. \& {Helmi}, A. 2024, \aap, 691, A277

\bibitem[{{Xue} {et~al.}(2008){Xue}, {Rix}, {Zhao}, {Re Fiorentin}, {Naab}, {Steinmetz}, {van den Bosch}, {Beers}, {Lee}, {Bell}, {Rockosi}, {Yanny}, {Newberg}, {Wilhelm}, {Kang}, {Smith}, \& {Schneider}}]{Xue2008ApJ...684.1143X}
{Xue}, X.~X., {Rix}, H.~W., {Zhao}, G., {et~al.} 2008, \apj, 684, 1143

\bibitem[{{Yang} {et~al.}(2019){Yang}, {Xue}, {Li}, {Zhang}, {Liu}, {Zhao}, {Chang}, {Tian}, \& {Li}}]{Yang2019ApJ...880...65Y}
{Yang}, C., {Xue}, X.-X., {Li}, J., {et~al.} 2019, \apj, 880, 65

\bibitem[{{Yang} {et~al.}(2022){Yang}, {Zhu}, {Tahmasebzadeh}, {Xue}, \& {Liu}}]{Yang2022AJ....164..241Y}
{Yang}, C., {Zhu}, L., {Tahmasebzadeh}, B., {Xue}, X.-X., \& {Liu}, C. 2022, \aj, 164, 241

\bibitem[{{Zhang} {et~al.}(2025){Zhang}, {Xue}, {Zhu}, {Zhang}, {Yang}, {Shao}, {Chang}, {Wang}, {Tian}, {Zhao}, \& {Liu}}]{Zhang2025}
{Zhang}, L., {Xue}, X.-X., {Zhu}, L., {et~al.} 2025, arXiv e-prints, arXiv:2505.09243

\bibitem[{{Zhu} {et~al.}(2018){Zhu}, {van den Bosch}, {van de Ven}, {Lyubenova}, {Falc{\'o}n-Barroso}, {Meidt}, {Martig}, {Shen}, {Li}, {Yildirim}, {Walcher}, \& {Sanchez}}]{Zhu2018a}
{Zhu}, L., {van den Bosch}, R., {van de Ven}, G., {et~al.} 2018, \mnras, 473, 3000

\end{thebibliography}

\begin{appendix}
\section{The recovery of DM distributions for Auriga 25}
Auriga 25 has a recent minor merger which significantly pertubated its stellar disk and inner halo. The infall orbit and mass of this merger are similar to those of the LMC in the Milky Way, but its pertubation to the disk and inner halo are significantly stronger than the LMC did to the Milky Way.

\begin{figure}
\centering\includegraphics[width=15cm]{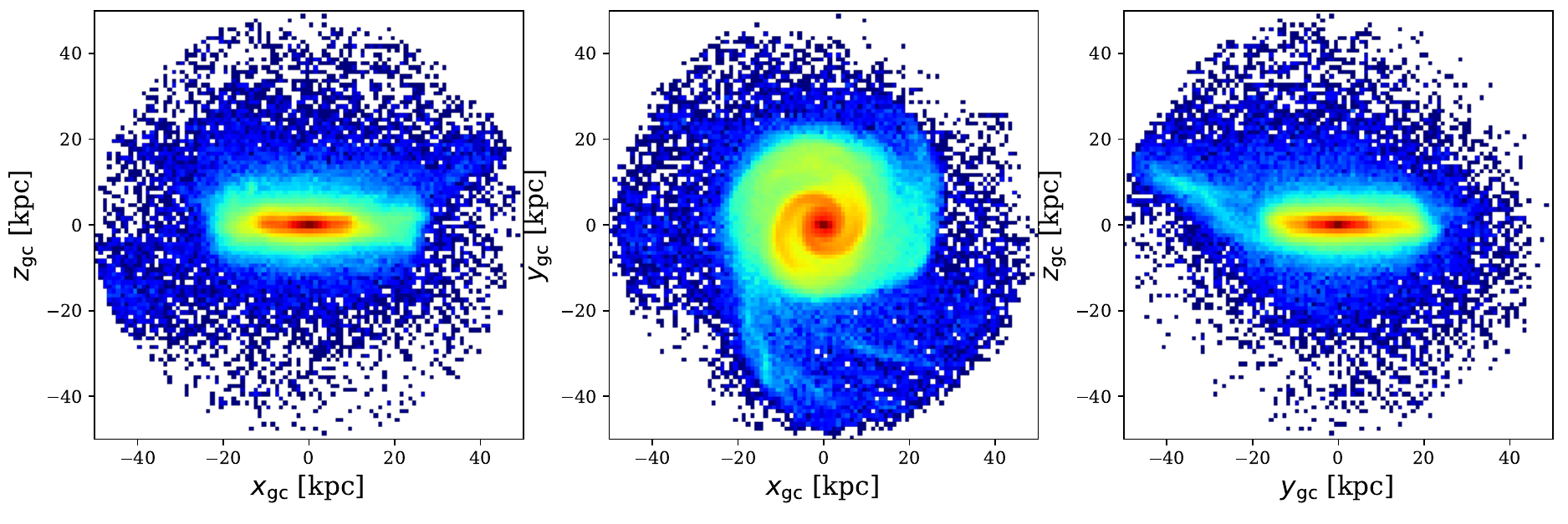}
\caption{The surface brightness of Auriga 25 projected with different directions. Its disk was significantly destroyed by a recent merger. Both its stellar disk and its halo are not in fully dynamical equilibrium.
}
\label{fig:SB_Au25}
\end{figure}

We orient Auriga 25 in the way that the LMC-like satellite is located in the southern hemisphere, and we take the particles in the north to create the mock data; the stars in the north look closer to dynamical equilibrium with less asymmetrical structures. The mock data are created in a similar way as for the other three galaxies. As shown in Figure~\ref{fig:SB_Au25}, there are significant substructures in the inner halo. We removed the big tail at the end of the disk, but not the other substructures.

\begin{figure}
\centering\includegraphics[width=10cm]{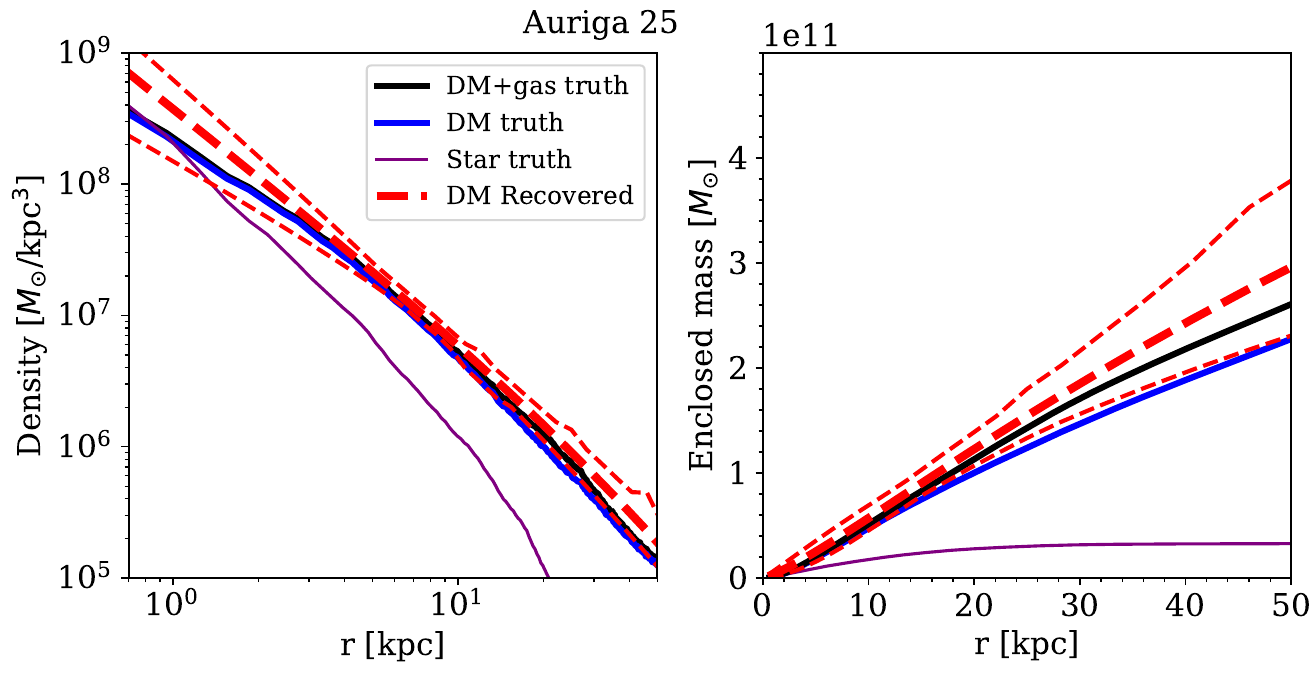}\centering\includegraphics[width=5.3cm]{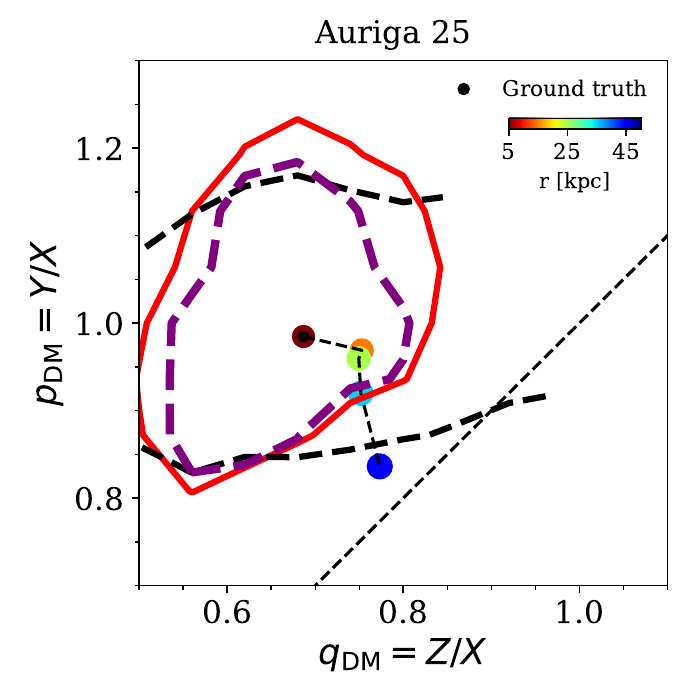}
\caption{The recovery of DM radial density distribution and DM 3D shape of Auriga 25, similar to that shown in Figure~\ref{fig:mass} and Figure~\ref{fig:pq3} for other galaxies.
}
\label{fig:pq_Au25}
\end{figure}

We applied the dynamical model to this galaxy in the same way as we did to the three galaxies we showed in the paper. In Figure~\ref{fig:pq_Au25}, we show the recovery of the DM radial density distribution and the 3D shape of the DM halo. In this case, we have larger statistical errors in the results. However, the true DM distribution is still generally recovered by our model within the $1\sigma$ confidence level. 

\end{appendix}

\end{document}